\documentclass[aps,prd,groupedaddress,superscriptaddress,nofootinbib]{revtex4}%
\usepackage{graphics}%
\usepackage[mathcal]{euscript}
\usepackage[latin1]{inputenc}
\usepackage{latexsym}
\usepackage{hyperref}
\usepackage{amsmath}    
\usepackage{graphicx}   
\usepackage{verbatim}  
\usepackage{color}      
\usepackage{subfigure}  
\usepackage{hyperref}   
\usepackage{bm}
\usepackage{amssymb}
\usepackage{bbm}
\usepackage{float}
\usepackage{slashed}
\usepackage{amsfonts}
\usepackage{euscript}
\usepackage{amsmath}    
\usepackage{mathrsfs} 
\usepackage{bbding}
\usepackage{pifont}
\usepackage{cancel}
\usepackage{siunitx}

\newcommand{\abs}[1]{\lvert#1\rvert}
\usepackage{booktabs}

\newcommand{\SubItem}[1]{{\setlength\itemindent{20pt} \item[-] #1}} 
\let\revappendix\appendix
\usepackage[capitalize]{cleveref}   

\begin{document}

\title{Electromagnetic tests of horizonless rotating black hole mimickers}

\author{Anna Zulianello}
\affiliation{Dipartimento di Fisica, Universit\`a di Trieste, Via A. Valerio 2, 34127 Trieste, Italy}
\author{Ra\'ul Carballo-Rubio}
\affiliation{Florida Space Institute, University of Central Florida, 12354 Research Parkway, Partnership 1, Orlando, FL, USA}
\author{Stefano Liberati}
\affiliation{SISSA, International School for Advanced Studies, Via Bonomea 265, 34136 Trieste, Italy}
\affiliation{INFN Sezione di Trieste, Via Valerio 2, 34127 Trieste, Italy}
\affiliation{IFPU, Institute for Fundamental Physics of the Universe, Via Beirut 2, 34014 Trieste, Italy}
\author{Stefano Ansoldi}
\affiliation{Dipartimento di Fisica, Università di Trieste, Via A. Valerio 2, 34127 Trieste, Italy}
\affiliation{IFPU, Institute for Fundamental Physics of the Universe, Via Beirut 2, 34014 Trieste, Italy}

\begin{abstract}

The interest in the implications that astrophysical observations have for the understanding of the structure of black holes has grown since the first detection of gravitational waves. Many arguments that are put forward in order to constraint alternative black hole models rely on substantial assumptions such as perfect spherical symmetry, which implies absence of rotation. However, given that astrophysical black holes will generally exhibit nonzero angular momentum, realistic constraints must take into account the effects of rotation. In this work we analyze the gravitational effect that rotation has on the emission from the surface of ultracompact objects, by studying how angular momentum affects the propagation of light rays. This allows us to evaluate the reliability of the constraints derived for supermassive black holes (more specifically, Sagittarius A* and M87*) assuming lack of rotation, as presented in [Astrophys. J. \textbf{701}, 1357 (2009); Astrophys. J. \textbf{805}, 179 (2015)]. We find that for rapidly spinning objects rotation can significantly affect the escaping probability of photon emitted from the surface of the object, with a significant increase at the equatorial regions and a decrease at the poles with respect to the non-rotating case. For not so rapidly spinning black hole candidates like Sagittarius A*, such modifications do not affect significantly the present constraints, which are nevertheless weaker than originally supposed due to the relativistic lensing here considered and additional phenomenological parameters that describe basic processes such as absorption. However, taking into account the angular dependence of the superficial emission of rapidly spinning black hole mimickers will be necessary for future studies of objects like \emph{e.g.} M87*.

\end{abstract}

\maketitle

\tableofcontents

\section{Introduction}

Our knowledge about astrophysical black holes\footnote{Note that throughout the paper we use the adjective \emph{astrophysical} to distinguish between astronomical sources and the mathematical notion of a black hole as a solution to the Einstein field equations.} is expanding at a rapid pace. Just a few years ago, astronomical observations could only characterize the environment of supermassive black holes, at distances that were still far away from the expected location of their boundaries, where general relativity predicts the existence of trapping horizons\footnote{While this issue is almost always neglected in the literature, it is important to keep in mind that the notion of event horizon is not observable by its own definition \cite{Visser2014}. Hence, when talking about observations one must resort to quasi-local definitions of horizons (see \emph{e.g.} \cite{Gourgoulhon2008,Krishnan2013} for reviews).}. The resulting constraints that could be placed on deviations from general relativity were therefore quite coarse; see \cite{Carballo-Rubio2018b} for a recent discussion. As a consequence, theoretical studies of alternatives to black holes aimed at addressing known issues in the framework of general relativity (in \cite{Chapline2000,Mazur2004,Visser2003}, for instance) were widely seen as purely mathematical studies far from testable outcomes; this did not hinder the steady growth of the number of models available in the literature (see \cite{Cardoso2019} for a recent review). There were nevertheless some exceptions to this perspective \cite{Broderick2007}, being some of them \cite{Narayan1997,Narayan2002,McClintock2004,Narayan2008,Broderick2009,Broderick2015} the subject of this paper.

The advent of gravitational-wave astronomy \cite{Abbott2016} has led to a shift in perspective, as black hole candidates in different ranges of masses can be now explored. Improvements in sensitivity will allow to understand the spacetime structure closer and closer (although not arbitrarily close) to their horizons \cite{Cardoso2016}. The Event Horizon Telescope \cite{Akiyama2019} has also provided quantitative and qualitative improvements with respect to previous observations using electromagnetic waves. These advances allow broader tests of alternatives to black holes \cite{Carballo-Rubio2018b,Cardoso2019}, although these tests are limited by the lack of knowledge of certain key properties (including their dynamical formation mechanism), as well as the complexities associated with modeling realistic situations that include all the ingredients necessary to represent realistic astrophysical situations.

Our paper presents a contribution towards a more accurate modeling of alternative models, often known as black hole mimickers, in which astrophysical black holes are modeled as ultracompact objects with a physical surface. The compactness of these objects is one of the relevant parameters that control the observational features associated with the existence of a physical surface \cite{Carballo-Rubio2018b}. Physical intuition indicates that black hole mimickers that are sufficiently compact must have observational features that are arbitrarily close to that of proper black holes, which in particular implies that so far observations cannot distinguish between these possibilities \cite{Cardoso2019}.

In this paper we revisit a well-known argument \cite{Narayan1997,Narayan2002,McClintock2004,Narayan2008,Broderick2009,Broderick2015} that is often invoked when discussing the viability of such alternatives. This argument asserts that equilibrium conditions that are thought to be attained between the supermassive black holes at the center of galaxies such as ours and their accretion disks imply that the presence of a physical surface must be accompanied by a faint emission that is nevertheless incompatible with observations of Sgr A* and M87*.\footnote{Let us stress however that the original conclusions in these papers have been shown to be somewhat premature, as these studies were not taking into account the effect of gravitational lensing \cite{Abramowicz2002,Lu2017,Carballo-Rubio2018,Carballo-Rubio2018b,Cardoso2019}, which leads to weaker contraints \emph{e.g.} on the viable compactness of the observed black holes candidates.} However, the calculations in \cite{Narayan1997,Narayan2002,McClintock2004,Narayan2008,Broderick2009,Broderick2015} assumed lack of rotation, which makes not possible to apply directly these arguments to physical systems in which non-negligible angular momentum is present (including in particular Sgr A* and M87*). The main goal of this paper is updating these calculations in order to include angular momentum and evaluating whether the resulting constraints are substantially modified.

 In general, taking into account rotation when discussing alternatives to black holes involves a number of technical complications, some of them being associated with the lack of knowledge of the surrounding vacuum spacetime (while in spherical symmetry Birkhoff's theorem determines completely the geometry in the external vacuum region, the results cannot be extended in the presence of rotation). These complications, and their possible impact on the argument described in \cite{Narayan1997,Narayan2002,McClintock2004,Narayan2008,Broderick2009,Broderick2015}, are briefly discussed in Sec. \ref{sec:back}, where we also explain the assumptions that are necessary for our subsequent analysis, which is presented in Sec. \ref{sec:core}. In Sec. \ref{sec:SgrA*} we applied our analysis to the particular case of Sagittarius A*. Our conclusions are discussed in Sec. \ref{sec:con}.

\section{Background and assumptions} \label{sec:back}

The original argument in \cite{Narayan1997,Narayan2002,McClintock2004,Narayan2008,Broderick2009,Broderick2015} formulated for spherically symmetric situations, as well as possible extensions, has been recently reviewed in \cite{Carballo-Rubio2018b,Cardoso2019}. Hence, here we will only summarize the main points that are relevant for our discussion below.

The gist of the argument is that any horizonless alternative to a black hole will emit in response to an accretion a certain amount of light, which depends on the emission and reflection coefficients on its surface, on the amount and spectral properties of the incoming light, and on the effect that the gravitational field has on propagating light rays (which, in turn, depends on the position of the surface). If the emitted amount of light can be calculated reliably, then it could be compared with observations in order to constrain the reflection coefficient or the position of the surface. As shown in \cite{Narayan1997,Narayan2002,McClintock2004,Narayan2008,Broderick2009,Broderick2015}, the total amount of emitted light can be calculated in spherical symmetry under the following assumptions:
\begin{enumerate}
\item{\emph{Accretion disk in steady state:} in the presence of an accretion disk, it is reasonable to expect that the compound system of the disk and the black hole alternative will reach a steady state if the latter is also emitting radiation as long as the interaction time is long enough.}
\item{\emph{Thermality of the surface:} the spectral properties of black hole alternatives would be model-dependent. However, if their exterior surface is assumed to be in thermal equilibrium, then the number of parameters that determine the spectrum of the emitted light is reduced down to a single parameter describing the total power radiated.}
\end{enumerate}
In spherical symmetry, one can see explicitly that there is a trade-off between the two assumptions above, depending on the value of the radius of the black hole alternative (or, which is equivalent, its compactness). As discussed originally in \cite{Narayan1997,Narayan2002,McClintock2004,Narayan2008,Broderick2009,Broderick2015}, gravitational lensing increases the number of interactions between different points on the surface, through light that is emitted at a given point but does not escape, being lensed back to a different point; the more compact the object is, the stronger this lensing effect becomes. Hence, it is reasonable to assume that the surface of compact enough objects will display a thermal behavior. On the other hand, these authors failed to notice that lensing also has a side effect that goes against the steady state assumption: as gravitational lensing becomes stronger, a smaller fraction of light rays can escape from the surface. In other words, the compactness of the black hole alternative determines the strength of the interaction between the disk and the object so that, if the latter is compact enough, this interaction could be so weak that the time scale to reach the steady state would be huge. Taking this second consequence of lensing into account is indispensable in order to extract meaningful constraints on the parameters describing the black hole alternative \cite{Carballo-Rubio2018b,Cardoso2019}.

When considering rotation the situation is more subtle. For instance, it is not straightforward to determine the interplay among the two assumptions above, the well-known phenomenon of superradiance (see \emph{e.g.} \cite{Brito2015}), and the associated instabilities of ultracompact objects with an ergoregion \cite{Cardoso2007,Maggio2018} (note that the existence of horizonless rotating black hole mimickers is not ruled out by this instability \cite{Comins1978,Chirenti2008,Maggio2017}). Moreover, the spacetime around horizonless alternatives to black holes will generally differ from the Kerr solution, although it has been shown that similar results to the well-known no-hair theorems should hold for compact enough objects \cite{Raposo2018,Barcelo2019}. 

A definitive answer to these issues will probably be obtained only through numerical simulations describing the interaction between ultracompact objects and radiation. Given that there are several important obstacles in the path towards these simulations, for instance the lack of known dynamical theories leading to the formation of these ultracompact alternatives, here we take a different approach and analyze the kinematical effects of rotation as a first step towards understanding the role that angular momentum may have for the constraints proposed in \cite{Narayan1997,Narayan2002,McClintock2004,Narayan2008,Broderick2009,Broderick2015}; from this perspective, our work is a follow-up (and a generalization) of \cite{Ogasawara2019}.

 In practice, this means that we will assume that the Kerr solutions provides a reasonable approximation for the purpose of extracting the leading-order effects of rotation on the trajectories of light (which is supported by the results of \cite{Raposo2018,Barcelo2019}), while we will implicitly assume that the steady state and thermality assumptions described above still hold (as we will see, in the presence of rotation one would expect that the temperature on the surface has an angular dependence, although thermalization can be assumed in the case of slow rotation). Further scrutiny on the validity of these assumptions is deferred to future works, but we will still be able to understand whether the kinematical effects associated with angular momentum are enough to change the main constraints that can be obtained following the arguments in \cite{Narayan1997,Narayan2002,McClintock2004,Narayan2008,Broderick2009,Broderick2015,Abramowicz2002,Lu2017,Carballo-Rubio2018,Carballo-Rubio2018b,Cardoso2019}.

\section{Our analysis} \label{sec:core}

\subsection{Basic elements of the Kerr metric} \label{sec:kerr}
While we are missing the analogue of the Birkhoff's theorem for axisymmetric solutions, as discussed above and suggested by recent results regarding generalized no-hair theorems \cite{Raposo2018,Barcelo2019} it is reasonable to approximate the spacetime outside an ultracompact object very close to form a horizon as the one of a rotating GR black hole.\footnote{It is nevertheless interesting to point out that considering non-Kerr metrics (see e.g. the discussion and references in \cite{Bambi2011}) would generally lead to modifications of the emission probabilities calculated below which, depending of their size, might allow to formulate new tests of no-hair theorems.} 
In the following we will work with the Kerr metric \cite{Stephani2003,Wiltshire2009}, in Boyer-Lindquist coordinates. The explicit form of the line element is given by: 
\begin{equation}
\text{d}s^2 = - \frac{\Sigma\Delta}{A} \text{d}t^2+\frac{\Sigma}{\Delta}\text{d}r^2+\Sigma \text{d}\theta^2+\frac{A}{\Sigma} \sin^2\theta \left(\text{d}\varphi-\frac{a (r^2+a^2-\Delta)}{A}\text{d}t\right)^2,
\label{Kerr_metric}
\end{equation}
where $\varphi$ is the azimuthal angle, $t$ is the coordinate over which the metric is stationary, $r$ and $\theta$ have the same meaning that in spherical symmetry, and, as usual, we have defined the functions
\begin{align}
&\Sigma = r^2+a^2 \cos^2\theta,\nonumber\\
&\Delta=r^2-2Mr+a^2,\nonumber\\
&A=(r^2+a^2)^2-a^2\Delta \sin^2\theta.
\end{align}
As it is well-known, this metric depends on two parameters, mass $M$ and spin $a$. The spin parameter $a$ is related to the angular momentum $J$ with respect to the rotation axis by $a=J/M$. Without loss of generality, we assume that $a\geq 0$. Note that we use units in which $c=1$ and $G=1$.

For $M> a$ the function $\Delta$ has two zeroes
\begin{equation}
r_\pm=M\pm\sqrt{M^2-a^2},
\label{rH}
\end{equation}
which are coordinate singularities that correspond to the external and internal horizon, respectively. In addition to the horizons, Kerr metric has also outer and inner ergospheres, determined by the condition that on them the norm of the Killing vector $\xi=\partial_t$ vanishes (in the following, we will denote by $\psi$ the Killing vector $\partial_\varphi$).
We will use the term ergosphere referring to the outer ergosphere $r_{\rm E}^+$:
\begin{equation}
r_{\rm E}^{\pm}=M\pm \sqrt{M^2-a^2 \cos^2\theta}.
\end{equation}
that is tangent to the outer horizon $r_+$ at  $\theta=0,\pi$ but it lies outside the horizon for other values of $\theta$. Thus, the ergoregion is defined by $r_+<r<r_{\rm E}^+$ and it corresponds to the region where $\xi$ can become spacelike.\\
As said, our motivation to analyze the Kerr spacetime is that it should provide a reasonable approximation for the external spacetime surrounding a rotating ultracompact object \cite{Raposo2018,Barcelo2019}, without a trapping horizon. In fact, we will assume that spacetime is not vacuum inside a region defined by $r\leq r_\star$, where $r_\star=r_+(1+\mu)$. The surface of the ultracompact object can be arbitrarily close to $r_+$ as $\mu\rightarrow0$. This is the natural generalization of analyses assuming spherical symmetry \cite{Narayan1997,Narayan2002,McClintock2004,Narayan2008,Broderick2009,Broderick2015,Abramowicz2002,Lu2017,Carballo-Rubio2018,Carballo-Rubio2018b,Cardoso2019}.\\

The surface at $r=r_\star$ will generally have an ellipsoidal shape.
Let us now consider the set of two-dimensional surfaces formed by fixing the $t$ and $r$ coordinate, with the last taking the specified fixed value $r=r_\star$, and letting the other two angular coordinates ($\theta$,$\varphi$) run over their respective ranges.
On these two-dimensional surfaces the induced metric results to be

\begin{equation}
\text{ds}^2_{\small \text{2-surface}} = \Sigma \text{d}\theta^2+\frac{A}{\Sigma} \sin^2\theta \text{d} \varphi^2,
\end{equation}

We need to calculate the determinant of this 2-metric, which is given by
\begin{equation}
\tilde{g} = \text{det}(g_{ij})_{\small \text{2-surface}} = A \sin^2 \theta = \left[(r^2+a^2)^2-a^2\Delta \sin^2\theta\right] \sin^2 \theta.
\label{inducedmetric}
\end{equation}

From the square root of Eq. \eqref{inducedmetric} we can obtain the surface element of the ellipsoid at $r=r_\star$:
 \begin{equation}
 \text{d}S = \sqrt{\tilde{g}} \ \text{d} \theta \text{d} \varphi = \sqrt{A} \sin \theta \ \text{d} \theta \text{d} \varphi
 \label{surf}
 \end{equation}
 and also their total surface:
 \begin{equation}
 S= \int{ \sqrt{A} \sin \theta \ \text{d} \theta \text{d} \varphi}
 \label{S_K}
 \end{equation}
 which will be useful for the discussion in the following sections.
 
\subsection{Geodesic equations and escape of a photon to infinity}
In order to analyze the role of the aforementioned lensing effects, let us review the equations for null geodesics in the Kerr spacetime. Let $k ^{\mu}$ be the 4-momentum of a photon, using the Hamilton-Jacobi method as discussed in \cite{Carter1968} it is possible to derive its components as
\begin{align}
 & k^t = \frac{1}{\Sigma}\left[a (L-aE \sin^2\theta)+\frac{a^2+r^2}{\Delta}[(r^2+a^2)E-aL]\right], \nonumber\\
& k^r = \frac{\sigma_r}{\Sigma} \sqrt{R}, \nonumber\\
& k^{\theta} = \frac{\sigma_{\theta}}{\Sigma} \sqrt{\Theta}, \nonumber\\
& k^{\varphi} = \frac{1}{\Sigma} \left[\frac{L}{\sin^2\theta}-aE+\frac{a}{\Delta}[(r^2+a^2)E-aL]\right], \label{eq4} 
 \end{align}
where $\sigma_r$, $\sigma_{\theta} = \pm$ and 
\begin{align}
& R = \left[(r^2+a^2)E-aL\right]^2-\Delta \left[(L-aE)^2+Q\right],\label{eq5}\\
& \Theta = Q-\cos^2\theta \left[\frac{L^2}{\sin^2\theta}-a^2E^2\right]. \label{eq6}
\end{align}
Here $E=-\xi^{\mu} k_{\mu}$, $L = \psi^{\mu} k_{\mu}$, and $Q$ are respectively the conserved energy, angular momentum, and Carter constant along the geodesic. $R$ and $\Theta$ have been chosen so as to separate the radial equation from the angular one. Indeed, $R$ represents the function governing the motion in the $r$ direction and $\Theta$ represents the function governing the motion of the $\theta$ coordinate and it is related to the angular polar component of the photon momentum. \\
Let us now introduce the locally non-rotating frame (LNRF) that is a tetrad basis associated with observers who co-rotate with the background spacetime. The basis one-forms are given by
\begin{align}
& e^{0} = \sqrt{\frac{\Sigma \Delta}{A}} \text{d}t,\nonumber \\ 
& e^{1} = \sqrt{\frac{\Sigma}{\Delta}} \text{d}r,\nonumber \\ 
& e^{2} = \sqrt{\Sigma}\, \text{d}\theta,\nonumber \\
& e^{3} = \sqrt{\frac{A}{\Sigma}}\sin\theta\,\text{d}\varphi-\frac{a(r^2+a^2-\Delta)\sin\theta}{\sqrt{\Sigma A}}\text{d}t.
\end{align}
and satisfy $g_{\mu \nu}=\eta_{ab} e_{\mu}^{a} e_{\nu}^{b}$, where $\eta_{ab}=\mbox{diag} (-1,1,1,1)$, and the $a$, $b$ indices run from 0 to 3. Using Eq. \eqref{eq4}, the tetrad components of the 4-momentum, $k^{a}=e_{\mu}^{a} k^{\mu}$, are then given by
\begin{align}
& k^{0}=\sqrt{\frac{\Delta}{\Sigma A}}\left[a(L-aE \sin^2\theta)+\frac{r^2+a^2}{\Delta}[(r^2+a^2)E-aL]\right],\nonumber  \\
& k^{1}= \sigma_r \sqrt{ \frac{R}{\Sigma \Delta}},\nonumber\\
& k^{2}= \sigma_{\theta} \sqrt{ \frac{\Theta}{\Sigma}},\nonumber \\
& k^{3}=\frac{L}{\sin\theta}\sqrt{\frac{\Sigma}{A}} \label{eq10}.
\end{align}
We can now follow the analysis in \cite{Ogasawara2019} in order to determine which light rays emitted at $r=r_\star$ escape at infinity,  and which ones are lensed back to the surface due to the strong gravitational fields.\\
We consider the emission of a photon near the surface $r=r_\star$ to infinity. We adopt units in which $M=1$ and assume that $a \neq 0$ in what follows\footnote{The case $a=0$ is considered as the limit case to be recovered for null rotation.}.\\
We study constant radius orbits as they can provide useful information to understand which photons can escape to infinity. Let us stress out that a constant-$r$ motion does not necessarily imply a constant-$\theta$ motion (since Eq. \eqref{eq5} is independent of the angular parameter and viceversa Eq. \eqref{eq6} is independent of the radial parameter), so motion at constant radius can take place on a spatially bidimensional hypersurface in $\theta$ and $\varphi$. Spherical orbits are orbits at constant radii living in these bidimensional hypersurfaces and that therefore are not necessarily confined to the equatorial plane. Such orbits represent a non-trivial generalization of the two circular orbits that instead lie in the equatorial plane \cite{Teo2003}. \\
We investigate the radial turning points to get an intuition of the region where photons with given energy and angular momentum are trapped, and cannot escape to infinity. In turn, this also allows us to understand which photons can be seen by a distant observer.\\

Let us introduce the dimensionless parameters
\begin{equation}
b=\frac{L}{E}, \ q=\frac{Q}{E^2}
\end{equation}
and

\begin{align}
&\hat{R}=\frac{R}{E^2} =[(r^2+a^2)-ab]^2-\Delta [(b-a)^2+q],\label{eqA} \\
&\hat{\Theta}=\frac{\Theta}{E^2}= q- \cos^2\theta\left[\frac{b^2}{\sin^2\theta}-a^2\right],\label{eqB} \\
\nonumber
\end{align}
for $E>0$ and, since we are looking for spherical orbits with constant radius $r$, we can impose the condition $\hat{R}(r)=0$.
Solving this for $b$, we obtain 
\begin{equation} \begin{split}
b=b_{1,2}(r)= \frac{-2 a r \pm \sqrt{\Delta [r^4-q(\Delta-a^2)]}}{\Delta-a^2}
\end{split} \end{equation}
We focus on extremum points of $b_{1,2}$. The positions are determined by the equations
\begin{equation}
b'_i(r)=0 \ , \ i = 1, 2
\end{equation}
Solving these for $q$, we obtain two classes of solutions parameterized in terms of $r$: 
\begin{align}
&\text{Class\  (i)}
\begin{cases}
\displaystyle b=\frac{r^2+a^2}{a} \\
\displaystyle q=-\frac{r^4}{a^2}\label{eq1_2}\\
\end{cases}\\
&\text{Class\  (ii)} 
\begin{cases}
 \displaystyle b= - \frac{r^3-3r^2+a^2r+a^2}{a(r-1)} \\
 \displaystyle q=-\frac{r^3 (r^3-6r^2+9r-4a^2)}{a^2 (r-1)^2}\label{eq1_1}
\end{cases}
\end{align}
Let us anticipate that indeed only one of these classes will be physically relevant given that, as we shall see below, Class(i) does not admit real values of $q$ for which $\hat{\Theta}$ is non negative as required by Eq. \ref{eq10}.

If we set $u=\cos\theta$, then Eq. \eqref{eqB} can be rewritten as 
\begin{equation}
\left(\frac{\Sigma}{E}\right)^2 \dot{u}^2 = \tilde{\Theta}(u)= q-(q+b^2-a^2)u^2-a^2u^4
\label{eq11}
\end{equation}

This condition gives us the physically allowed ranges for $u$ that can be easily found imposing $\tilde{\Theta}(u)=0$ and solving the resulting quadratic equation in $u^2$.
Three cases should be distinguished according to the sign of $q$.

When $q>0$, a situation that can only be realized in Class (ii), there is only one positive root and it is given by  
\begin{equation}
u_0^2=\frac{(a^2-q-b^2)+\sqrt{(a^2-q-b^2)^2+4a^2q}}{2a^2}.
\end{equation}
The physically allowed range for $u$ in this case is between $\pm \abs{u_0}$, meaning that such orbits cross the equatorial plane repeatedly at points that are referred to as \emph{nodes} of the orbit. 
This behavior corresponds to a radius lying in the range $r_{1,{\rm min}} \leq r \leq r_{2,{\rm min}}$, where $r_{1,{\rm min}}$ and $r_{2,{\rm min}}$ are the two radii at which $q=q_{\rm min}$ and $q_{\rm min}$ is the value for which Eq. \eqref{eqB} vanishes.

When $q<0$, the right-hand side of Eq. \eqref{eqB} is non negative only if
\begin{equation}
a^2-q-b^2>0
\label{12}
\end{equation}
holds.

It turns out that Class (i) can immediately be ruled out in this case, since by Eq. \ref{eq1_2},  $a^2-q-b^2=-2r^2<0$. For Class (ii) $q$ can be negative but note that, in general, one can show that
\begin{equation} 
a^2-q-b^2= -\frac{2r(r^3-3r+2a^2)}{(r-1)^2} < 0\\
\label{eqU}
\end{equation}

for any $r \geq r_+ \geq 1$. This can be done by proving that $r^3-3r+2a^2 \geq 0$. Let us then consider $r=r_++\delta$ with $\delta \geq 0$, we can write

\begin{equation}
\begin{split}
r^3-3r+2a^2 &= (r_++\delta)^3-3(r_++\delta)+2a^2\\
&= r_+^3+3r_+^2 \delta+3 r_+ \delta^2 +\delta^3 -3r_+-3\delta+2a^2\\
&=r_+^3+3\delta(r_+^2-1)+3 r_+ \delta^2+\delta^3-3 r_++2a^2\\
&\geq r_+^3-3r_++2a^2\\
&=(1-a^2)(1+\sqrt{1-a^2})\geq 0\\
\end{split}
\end{equation}
So the case $q<0$ is irrelevant for us, in the sense that there are no spherical orbits with $q<0$ (in other words, there are no turning points for these trajectories). This does not imply that there are no escaping trajectories with $q<0$, which are naturally included in our discussion below.

When $q=0$, the two roots are $u_0^2=0$ and $1-b^2/a^2$. These cases describe equatorial orbits: for $b^2\geq a^2$, there is only one relevant root ($u_0^2=0$), but for $b^2<a^2$ they are both relevant. \\ 

\begin{figure}[b!]
  \centering
  \includegraphics[width=8 cm, height=6.5 cm]{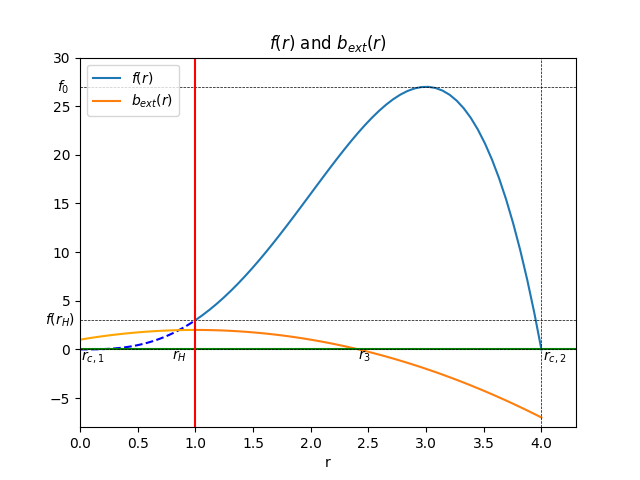}
  \includegraphics[width=8 cm, height=6.5 cm]{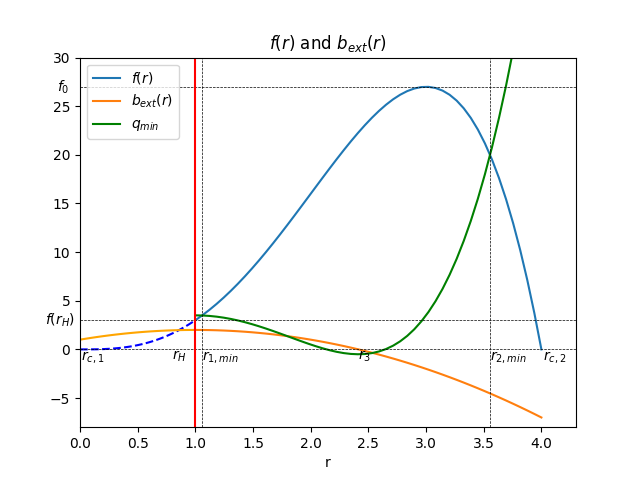}
  
  \caption{For $a=1$ typical numerical plots of $f(r)$ (blue solid and dashed lines), $b(r)$ (orange lines) and $q_{\rm min}$ (green lines) are shown. The blue solid lines show $f(r)$ in the range $r>r_+$ and $f(r_+)<q<q_{\rm max}$ for the case of $\theta=\pi/2$ (left) and in the range $r>r_{1,{\rm min}}$ and $q_{1,{\rm min}}<q<q_{\rm max}$ for a case away from the equatorial plane ($\theta=\pi/4$)(right).
  }
  \label{fig: qb}
\end{figure}

\begin{figure}[b!]
  \centering
  \includegraphics[width=8 cm, height=6.5 cm]{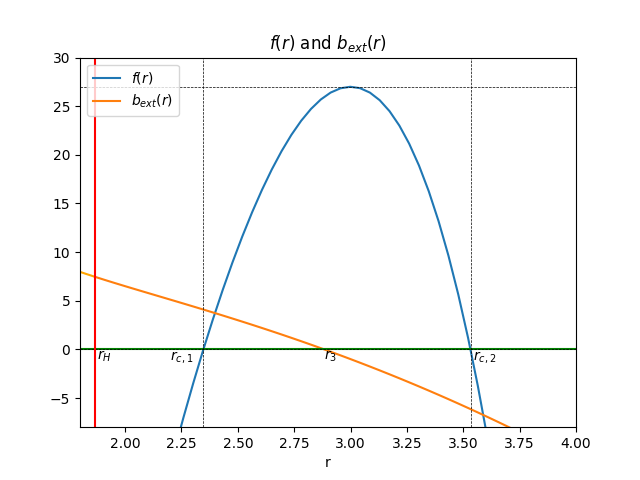}
  \includegraphics[width=8 cm, height=6.5 cm]{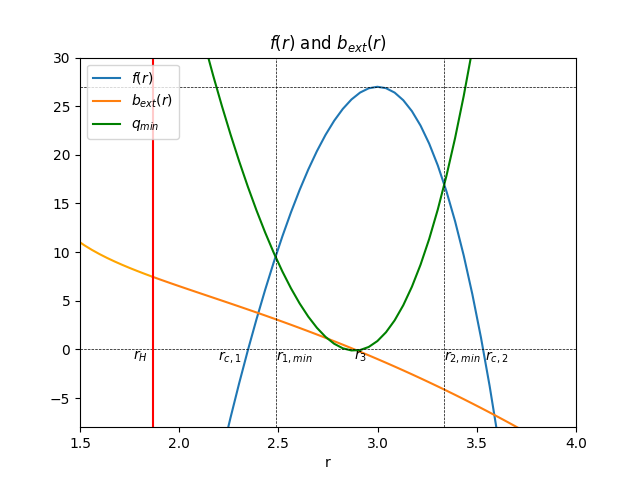}
  
  \caption{For $0<a<1$ typical numerical plots of $f(r)$ (blue solid and dashed lines), $b(r)$ (orange lines) and $q_{\rm min}$ (green lines) are shown. The blue solid lines show $f(r)$ in the range $r>r_{c,1}$ and $f(r_{c,1})<q<q_{\rm max}$ for the case of $\theta=\pi/2$ (left) and in the range $r>r_{1,{\rm min}}$ and $q_{1,{\rm min}}<q<q_{\rm max}$ for a case away from equatorial plane ($\theta=\pi/4$)(right). 
  }
  \label{fig: qb05}
\end{figure}

Given the above analysis we shall now focus on the properties of Class (ii). Let us then start from Eq. \eqref{eq1_1} expressing $q$ as a function of $r$
\begin{equation}
 q=f(r)=-\frac{r^3 (r^3-6r^2+9r-4a^2)}{a^2 (r-1)^2}
 \label{eq13}
 \end{equation}
As said, for positive $q$, one gets that $q$ is bounded from below with $q_{\rm min}$ determined by the vanishing of the quantity in Eq. \eqref{eqB}. So that
\begin{equation}
q \geq q_{\rm min}=\cos^2\theta \left[ \frac{b^2}{\sin^2\theta}-a^2\right].
\end{equation}
More generally, $q$ satisfy the inequalities $\min(q_{1,{\rm min}}, q_{2,{\rm min}} ) \leq q_{\rm max}$, where $q_{1,{\rm min}}$ and $q_{2,{\rm min}}$ (and the corresponding $r_{1,{\rm min}}$ and $r_{2,{\rm min}}$) are given by $f (r) = q _{\mathrm{min}}$.
From Eq. \eqref{eq5}, instead, we obtain a condition on the maximum of $q$, that always corresponds to a fixed value, $q_{\rm max}= 27$ for $r=r_{\rm max}=3$, as we can see in Figs. \ref{fig: qb} - \ref{fig: qb05}.


At the would-be horizon, $f$ takes the value 
\[
f(r_+)=
\begin{cases}
 4-1/a^2=3 & (a=1) \\
 -r_+^4/a^2 & (0<a<1)\\
\end{cases}
\]
which is always positive only for $a=1$.

Two cases can be defined according to the value of $f(r_+)$ and they correspond to $a=1$ and $0<a<1$. In principle we could also consider the case $a>1$, as in a black hole mimicker we expect the interior region to deviate from the Kerr solution, and we could thus avoid the pathologies of the Kerr metric. However, the existence of the Thorne limit \cite{Thorne1974}, suggests that it may not be easy to spin-up objects beyond the $a=1$, and, therefore, we choose to restrict our attention to $a \leq 1$ in this work.
Moreover, the case $a=0$ is the Schwarzschild limit, which in this discussion is considered only as the limiting case to be recovered in the limit of null (very slow) rotation.

The classification introduced above enables us to investigate the roots of Eq. \eqref{eq13} outside the horizon, \emph{i.e.} the radii of spherical photon orbits. \\
We shall describe the possible cases as:
\begin{itemize}
\item Case $a=1$: It has some peculiarities because for this case the position of $r_{1,{\rm min}}$ changes with respect to the position of the horizon and so it is convenient to divide it into three different cases depending on the values of $\theta$ w.r.t. $\theta_+$ where $\theta_+ \simeq \ang{47}$ has been obtained imposing the necessary condition that $\hat{\Theta} \geq 0$ in $r_+$.
 \SubItem{$\theta > \theta_+$}: Eq. \eqref{eq13} in the range $f(r_+)<q \leq q_{\rm max}$ has two roots $r_i(q)$  ($r_1 \leq r_2$) outside the horizon. On the other hand, the equation in the range $0 \leq q \leq f(r_+)$ only has the largest root $r_2(q)$ outside the horizon, while the root $r_1(q)$ lies inside the horizon because $q_{1,{\rm min}}< f(r_+)$;
\SubItem{$\theta = \theta_+$}: it corresponds to $q_{1,{\rm min}}=f(r_+)$. Eq. \eqref{eq13} in this case has one of the two roots, $r_1(q)$  with $r_1 < r_2$, exactly on the horizon ;
\SubItem{$\theta < \theta_+$}: it corresponds to the condition $q_{1,{\rm min}}>f(r_+)$. Eq. \eqref{eq13} in the range $q_{1,{\rm min}}<q \leq q_{\rm max}$ has two roots $r_i(q)$  ($r_1 \leq r_2$) outside the horizon; \\
\item Case $0<a<1$:
Eq. \eqref{eq13} in the range $q_{1,{\rm min}} \leq q \leq q_{\rm max}$ has the two roots $r_i(q)$ ($r_1 \leq r_2$) outside the horizon. \\
Notice that $r_i $ ($i=1,2$) coincide with $r_{\rm max}$ in the case $q=q_{\rm max}$. In particular, if $q=0$ ($\theta=\ang{90}$), the radii $r_i$ reduce to those of circular photon orbits 
\begin{equation}
r_{c,i} = r_i(q=0).
\end{equation}
\end{itemize}
\begin{table}[h]
\center
\small
\begin{ruledtabular}
\begin{tabular}{c|c|c|c|c}
Cases & $q$ & $\sigma_r=+$ & $\sigma_r=-$ & marginal pairs\\
 for $a=1$&&&& of ($\sigma_r,b$)\\
\hline
(a) : $r_1<r_+<r_\star$ & $q_{1,{\rm min}} \leq q <f(r_+)$ & $b_2^s<b \leq b_1(r_\star)$ & $b_1(r_+)<b<b_1(r_\star)$ & ($+,b_2^s$) and ($-,b_1(r_+)$)\\
&&&&\\
(b) : $r_+<r_1<r_\star$ & $f(r_+)<q<f(r_\star)$ & $b_2^s<b \leq b_1(r_\star)$ & $b_1^s<b<b_1(r_\star)$ & ($+,b_2^s$) and ($-,b_1^s$)\\
&&&&\\
(c) : $r_+<r_\star<r_1$ & $f(r_\star)<q<q_{\rm max}$ & $b_2^s<b \leq b_1^s$ & n/a  & ($+,b_2^s$) and ($+,b_1^s$)\\
\hline
\hline
Cases & $q$ & $\sigma_r=+$ & $\sigma_r=-$ & marginal pairs\\
 for $0<a<1$&&&& of ($\sigma_r,b$)\\
\hline
(b) : $r_+<r_1<r_\star$ &$ q_{1,{\rm min}}<q<f(r_\star)$& $b_2^s<b \leq b_1(r_\star)$ & $b_1^s<b<b_1(r_\star)$ & ($+,b_2^s$) and ($-,b_1^s$)\\
&&&&\\
(c) : $r_+<r_\star<r_1$ & $f(r_\star)<q<q_{\rm max}$ & $b_2^s<b \leq b_1^s$ & n/a  & ($+,b_2^s$) and ($+,b_1^s$)\\
\end{tabular}

\label{Tabella1}

\caption{Range of $b$ in which a photon can escape from $r=r_\star$ to infinity. The last column shows two pairs ($\sigma_r,b$) of the marginal parameter values with which a photon cannot escape to infinity for each cases.}
\end{ruledtabular}
\end{table}
\normalsize

The extremal values of $b_i$ become
\begin{equation}
b_i^s = b_i(r_i)|_{q=f(r_i)}= - \frac{r^3-3r^2+a^2r+a^2}{a(r-1)}
\end{equation}
which are values of the impact parameter of photons on spherical photon orbits.
We need to determine the range of $b$ in which a photon can escape from $r=r_\star$ to infinity. In order to do this, we define three cases according to the relative position of $r_1$ with respect to to $r_+$ and $r_\star$:
\begin{equation}
\begin{array}{ll}
$Case (a)$: & r_1 < r_+ < r_\star\\
$Case (b)$: & r_+ \leq r_1 < r_\star\\
$Case (c)$: & r_+ < r_\star \leq r_1
\end{array}
\label{15}
\end{equation}
where $r_\star$ is defined in the range $r_+ < r_\star \leq r_{\rm max}$. 

For Case (a), that appears only for $a=1$, as $r$ increases from $r_+$ to $\infty$, $b_1$ begins with $b_1(r_+)$, given by
\begin{equation}
b_1(r_+)=
\begin{cases}
\displaystyle 2 & (a=1) \\
\displaystyle \frac{2(1+\sqrt{1-a^2})}{a} & (0<a<1)\\
\end{cases}
\end{equation}
and monotonically increases to $\infty$. For Cases (b) and (c), as $r$ increases from $r_+$ to $\infty$, $b_1$ starts from $b_1(r_+)$, decreases to a local minimum $b_1^s$ at $r=r_1$, and then increases to $\infty$. 

We have not discussed explicitly trajectories with $q<0$, as these trajectories have no associated spherical orbits such as the ones occurring for $q>0$. This does not mean that we are not taking into account trajectories with $q<0$ that can escape. Given that $q_{\rm min}$ (let us recall that the latter is defined by the condition $\hat{\Theta}=0$) can become negative for certain values of $b$, our analysis includes automatically these escaping trajectories. This can also be seen graphically, for instance from the fact that the right panel in Fig. \ref{fig: qb} includes a region in which $q_{\rm min}$ is negative. However, we will see this even more explicitly in the calculation of escape probabilities below.

\subsection{Escape cone and critical angles}

The photon emission angles ($\alpha, \beta$) of a light ray with respect to the LNRF are defined as follows \cite{Ogasawara2019}: 
$$\begin{array}{ll}
k^{a}|_{r=r_\star} \propto (1, \cos\alpha \ \sin\beta, -\cos\beta, \sin\alpha \ \sin\beta),
\end{array}$$
where $\alpha\in[-\pi,\pi]$ and $\beta\in[0,\pi]$. 
Equivalently,
\begin{align}\label{eq:scab}
& \sin\alpha= \frac{k^{3}}{\sqrt{(k^{1})^2+(k^{3})^2}}\Bigg\vert_{r=r_\star} = \frac{b\Sigma\sqrt{\Delta}}{\sqrt{A\hat{R}\sin^2\theta+b^2\Sigma^2\Delta}}\Bigg\vert_{r=r_\star},\nonumber\\
& \cos\alpha= \frac{k^{1}}{\sqrt{(k^{1})^2+(k^{3})^2}}\Bigg\vert_{r=r_\star} = \frac{\sigma_{r}\sqrt{A\hat{R}}\ \sin\theta}{\sqrt{A\hat{R} \sin^2\theta+b^2\Sigma^2\Delta}}\Bigg\vert_{r=r_\star},\nonumber\\
& \sin\beta = \frac{\sqrt{(k^{1})^2+(k^{3})^2}}{\sqrt{(k^{1})^2+(k^{2})^2+(k^{3})^2}}\Bigg\vert_{r=r_\star} = \sqrt{\frac{\hat{R}A \sin^2\theta+b^2 \Sigma^2 \Delta}{(\hat{R}+\Delta  \hat{\Theta})A \sin^2\theta+b^2\Sigma^2\Delta}}\Bigg\vert_{r=r_\star},\nonumber\\
& \cos\beta = \frac{-k^{2}}{\sqrt{(k^{1})^2+(k^{2})^2+(k^{3})^2}} \Bigg\vert_{r=r_\star} = - \frac{\sigma_{\theta}\sqrt{\hat{\Theta}A\Delta}\sin\theta}{\sqrt{(\hat{R}+\Delta  \hat{\Theta})A \sin^2\theta+b^2\Sigma^2\Delta}}\Bigg\vert_{r=r_\star}.
\end{align}
Thus $\alpha$ is the angle between $e^{1} $ and $k$ projected onto the emission plane, which is positive in the direction $e^{3}$, and $\beta$ is the angle between $-e^{2}$ and $k$, where $k$ is the projection of $k^{\mu}$ normal to $e^{0}$. 

\begin{figure}[b!]
  \centering
  \includegraphics[width=8 cm, height=5.6 cm]{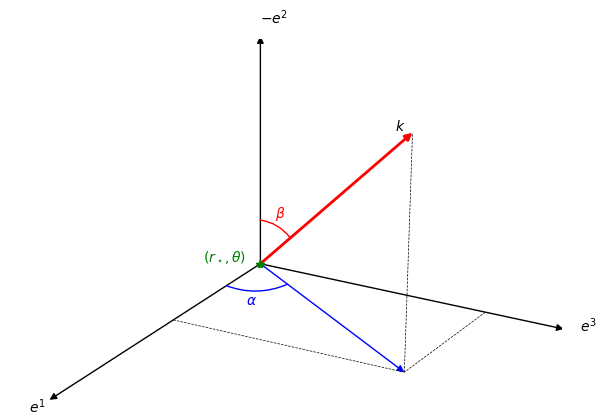}
  
  \caption{Emission angles ($\alpha,\beta$) in LNRF. The origin coincides with the emission point ($r_\star,\theta$).}
  \label{fig: em}
\end{figure}

We can define an escape cone $S$, in terms of $\alpha$ and $\beta$, as the solid angle of emission that allows photons to escape to infinity. Assuming that photons are emitted isotropically, we can define the escape probability $P(r_\star)$ by 
\begin{equation}
P(r_\star)= \frac{1}{4\pi} \int_{S} \text{d}\alpha \text{d}\beta \,\sin\beta.
\label{PP}
\end{equation}
This is the same definition provided in \cite{Ogasawara2019}, including the normalization factor of $1/4\pi$, which from a physical perspective is a consequence of considering photons emitted in all directions. As a consequence, our numerical results below are, whenever comparable (namely in the $a\rightarrow 1$ limit and $\theta=\pi/2$ case), equivalent to the results in \cite{Ogasawara2019} (see the tables in Appendix \ref{Tables}). It is worth mentioning that this implies that in the limit $a\rightarrow0$ there is a mismatch with the analytical results valid for spherical symmetry obtained in \cite{Carballo-Rubio2018b}, as in the latter only the photons emitted outwards where considered. Hence, our numerical results will differ in a multiplicative factor of $1/2$ with respect to these analytical results.

Given that we are interested in evaluating $P$, we need to determine the critical angles, marking the boundaries of the escape cone $S$.
There is a one-to-one correspondence between the critical angles and the parameter set ($\sigma_r, b, q$) for photons that cannot marginally escape to infinity.
Such parameters sets are summarized in the last column of Tab. \ref{Tabella1} and they represent the boundary of the ranges for photons that can escape to infinity. Therefore, they are defined as the marginal pairs (see \cite{Ogasawara2019}).

Finally we obtain the critical angles ($\alpha_i$, $\beta_i$) ($i=1,2$) relevant to marginal parameter values associated to $b_i$ and their total set $\partial S$ (\emph{i.e.} the boundary of $S$) as follows:
\begin{equation}
\partial S = \bigcup _{i=1,2} \biggl\{ (\alpha_i,\beta_i) \Big\vert {\rm min} \{q_{1,{\rm min}},q_{2,{\rm min}}\}  \leq q \leq q_{\rm max} \biggr\}
\end{equation}
where if $q_{1,min}<f(r_+)$ for $a=1$ we get

\begin{equation}
(\alpha_1,\beta_1)=
\begin{cases}
(\alpha_{1 (a)},\beta_{1 (a)})=(\alpha,\beta) \big\vert_{\sigma_r=-,b=b_1(r_+)} & \text{for} \ q_{1,{\rm min}} \leq q \leq f(r_+)  \ \text{[Case (a)]} \\
(\alpha_{1 (b)},\beta_{1 (b)})=(\alpha,\beta) \big\vert_{\sigma_r=-,b=b_1^s} & \text{for} \ f(r_+) \leq q \leq f(r_\star)  \ \text{[Case (b)]} \\
(\alpha_{1 (c)},\beta_{1 (c)})=(\alpha,\beta) \big\vert_{\sigma_r=+,b=b_1^s} &\text{for} \ f(r_\star) \leq q \leq q_{\rm max}  \ \text{[Case (c)]} 
\end{cases}
\end{equation}
if $q_{1,min}>f(r_+)$ always for $a=1$
\begin{equation}
(\alpha_1,\beta_1)=
\begin{cases}
(\alpha_{1 (b)},\beta_{1 (b)})=(\alpha,\beta) \big\vert_{\sigma_r=-,b=b_1^s} & \text{for}  \ q_{1,{\rm min}} \leq q \leq f(r_\star)  \ \text{[Case (b)]} \\
(\alpha_{1 (c)},\beta_{1 (c)})=(\alpha,\beta) \big\vert_{\sigma_r=+,b=b_1^s} &\text{for} \ f(r_\star) \leq q \leq q_{\rm max}  \ \text{[Case (c)]} 
\end{cases}
\end{equation}
for $0<a<1$
\begin{equation}
(\alpha_1,\beta_1)=
\begin{cases}
(\alpha_{1 (b)},\beta_{1 (b)})=(\alpha,\beta) \big\vert_{\sigma_r=-,b=b_1^s} & $for$ \ q_{1,{\rm min}} \leq q \leq f(r_\star) \ [$Case (b)$]\\
(\alpha_{1 (c)},\beta_{1 (c)})=(\alpha,\beta) \big\vert_{\sigma_r=+,b=b_1^s} & $for$\  f(r_\star) \leq q \leq q_{\rm max} \ [$Case (c)$]
\end{cases}
\end{equation}
 and finally for $ 0<a <1$ and $a=1$
\begin{eqnarray}
\begin{array}{lll}
(\alpha_{2},\beta_{2})=(\alpha,\beta) \big\vert_{\sigma_r=+,b=b_2^s} & $for$\ q_{2,{\rm min}} \leq q \leq q_{\rm max} & [$Cases (a) - (c)$]
\end{array}
\end{eqnarray}

 Note that once we fix the value of $r_\star$, then the critical angles ($\alpha_i$,$\beta_i$) depend only on $q$, \emph{i.e.}, $\alpha_i=\alpha_i(q)$ and $\beta_i=\beta_i(q)$.
In particular they depend only on the values of $r_i$ for which $f(r)=q$.

\begin{figure}[!htbp]
  \centering
  \includegraphics[width=6 cm, height=4.5 cm]{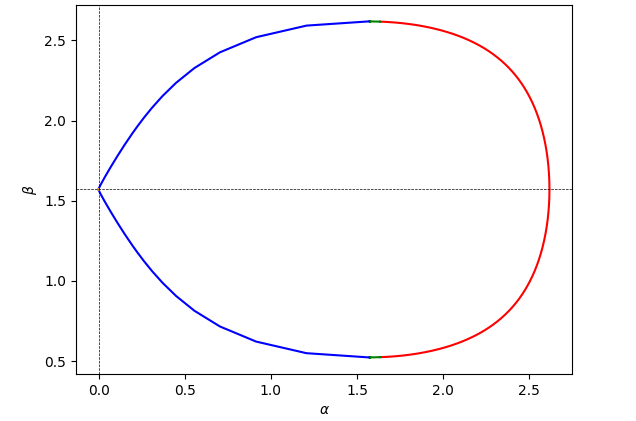}
  \includegraphics[width=6 cm, height=4.5 cm]{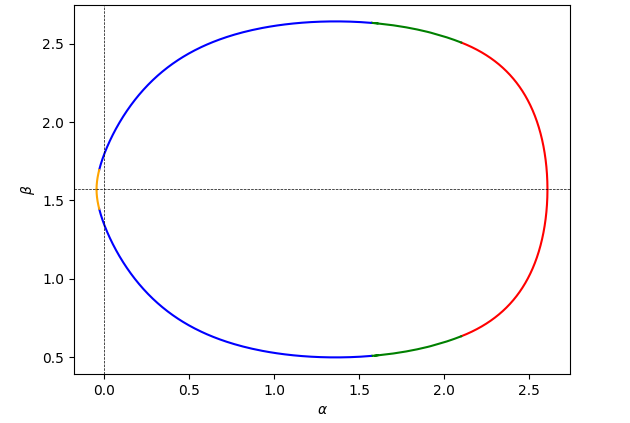}
  \includegraphics[width=6 cm, height=4.5 cm]{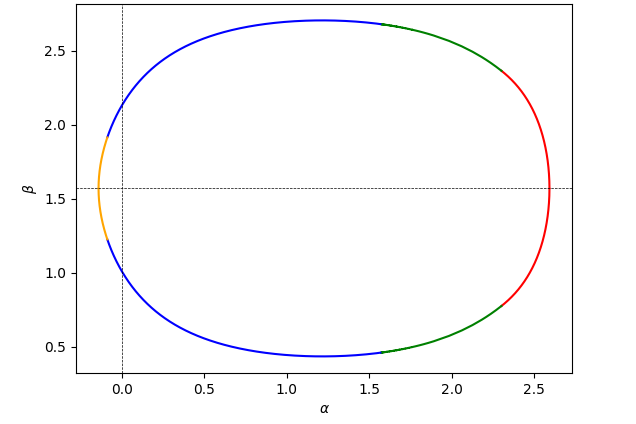}
   \includegraphics[width=6 cm, height=4.5 cm]{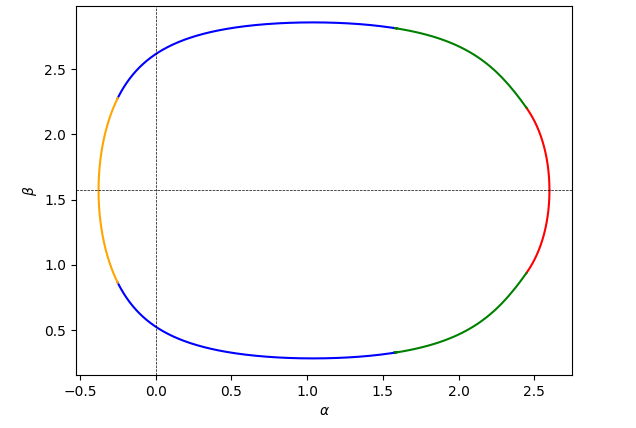}
   
  \caption{Critical angles in $\alpha-\beta$ plane in the extremal Kerr black hole ($a=1$). The red, green, blue and orange lines show ($\alpha_{1 (a)}, \beta_{1(a)}$), ($\alpha_{1 (b)}, \beta_{1(b)}$), ($\alpha_{1 (c)}, \beta_{1(c)}$) and ($\alpha_{2}, \beta_{2}$), respectively. We set $\mu=10^{-3},0.1,0.3,0.7$ in this order. Let us stress that the seemingly sharp transition in the first panel is actually smooth as the curves in the remaining panels, although zooming in is necessary in order to appreciate it.}
\label{fig: fig1}
\end{figure}

\begin{figure}[!htbp]
  \centering
  \includegraphics[width=4 cm, height=3.5 cm]{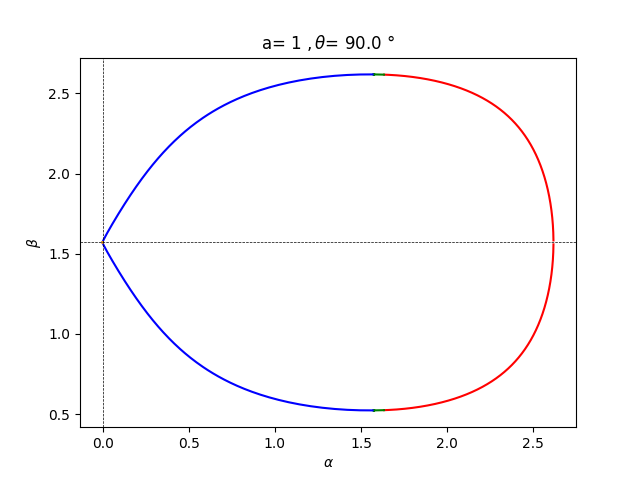}
  \includegraphics[width=4 cm, height=3.5 cm]{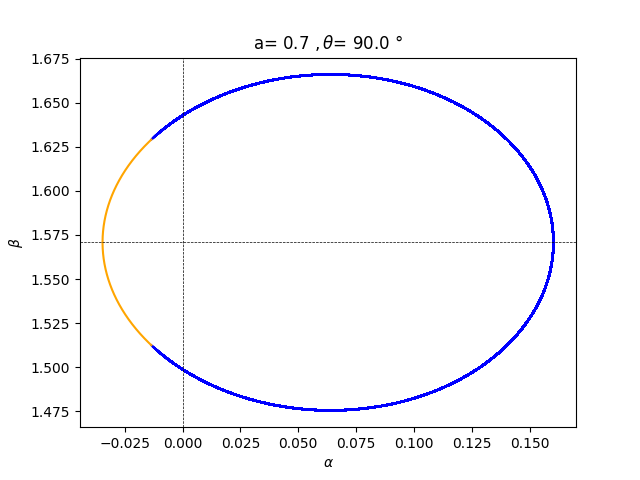}
  \includegraphics[width=4 cm, height=3.5 cm]{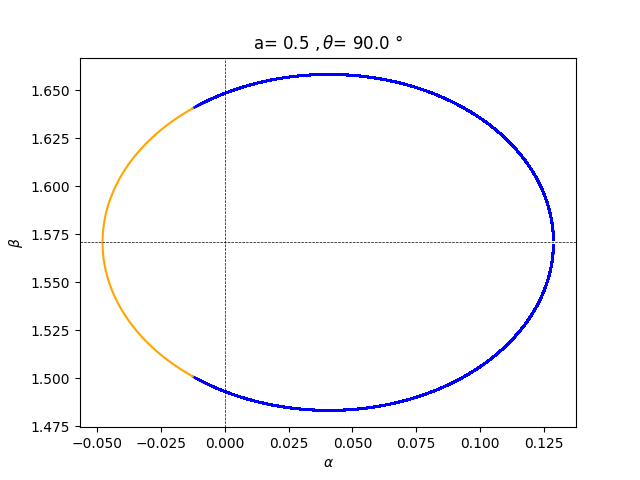}
   \includegraphics[width=4 cm, height=3.5 cm]{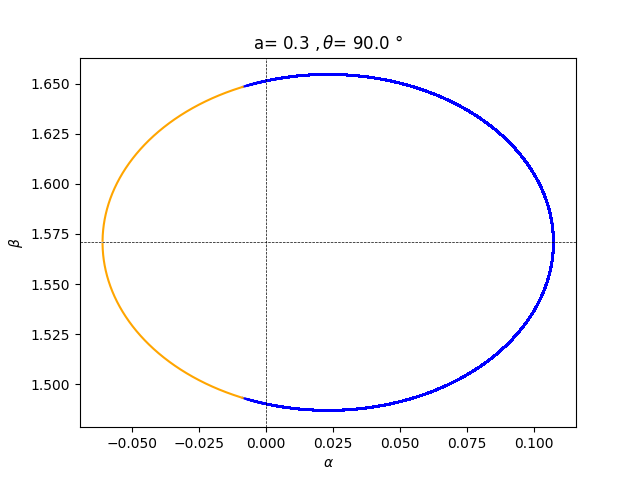}
   \includegraphics[width=4 cm, height=3.5 cm]{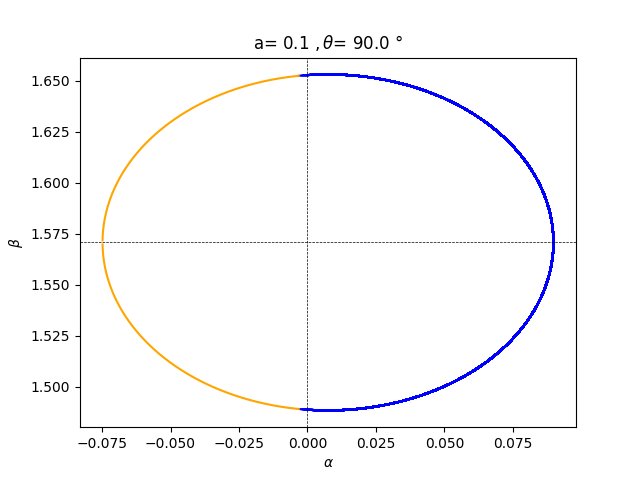}
    \includegraphics[width=4 cm, height=3.5 cm]{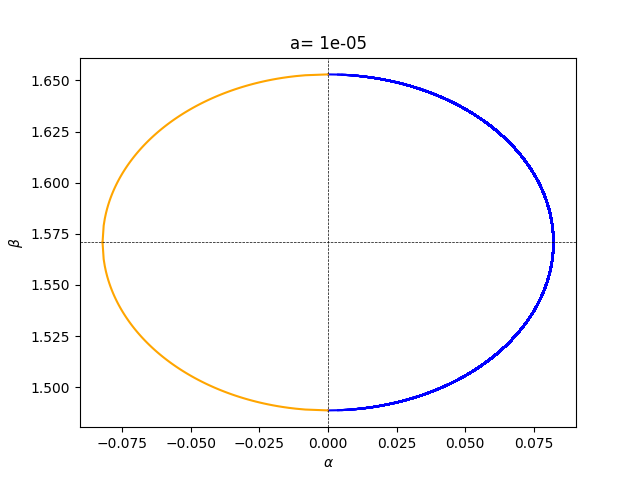}
    
  \caption{Critical angles in $\alpha-\beta$ plane in the Kerr black hole for different values of $a$ ($a=1,\ 0.7,\ 0.5,\ 0.3,\ 0.1,\ 10^{-5}$). The red, green, blue and orange lines show ($\alpha_{1 (a)}, \beta_{1(a)}$) (when it is present), ($\alpha_{1 (b)}, \beta_{1(b)}$), ($\alpha_{1 (c)}, \beta_{1(c)}$) and ($\alpha_{2}, \beta_{2}$), respectively. We set $\mu=10^{-3}$. Let us stress that the seemingly sharp transition in the first panel is actually smooth as the curves in the remaining panels, although zooming in is necessary in order to appreciate it.}
\label{fig: fig2}
\end{figure}

\begin{figure}[!htbp]
  \centering
  \includegraphics[width=4 cm, height=3.5 cm]{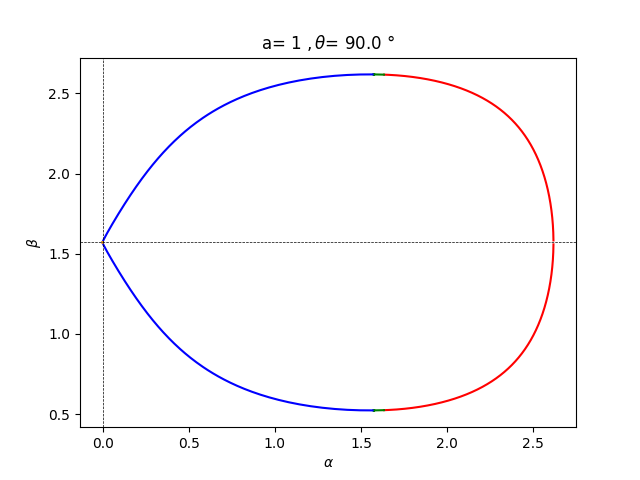}
  \includegraphics[width=4 cm, height=3.5 cm]{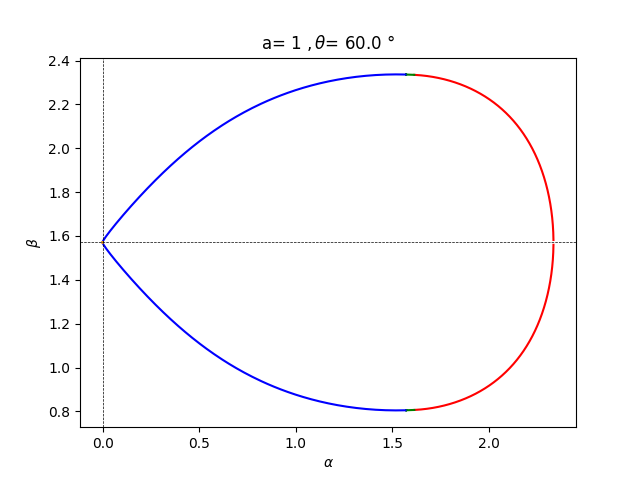}
  \includegraphics[width=4 cm, height=3.5 cm]{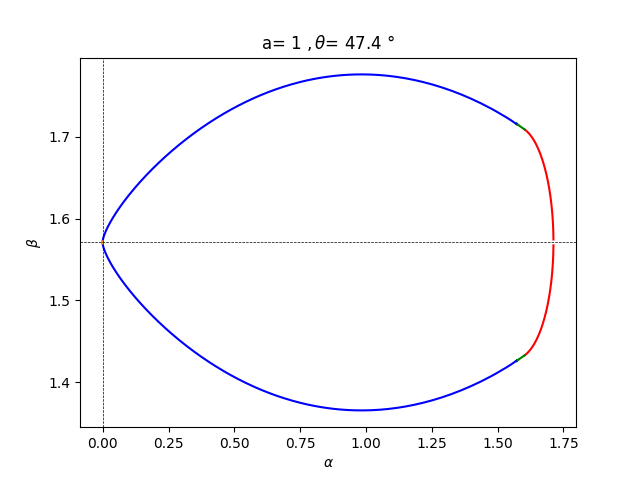}
   \includegraphics[width=4 cm, height=3.5 cm]{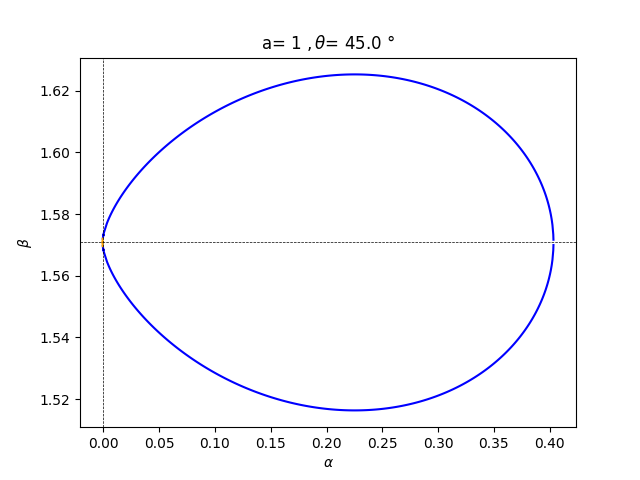}
   \includegraphics[width=4 cm, height=3.5 cm]{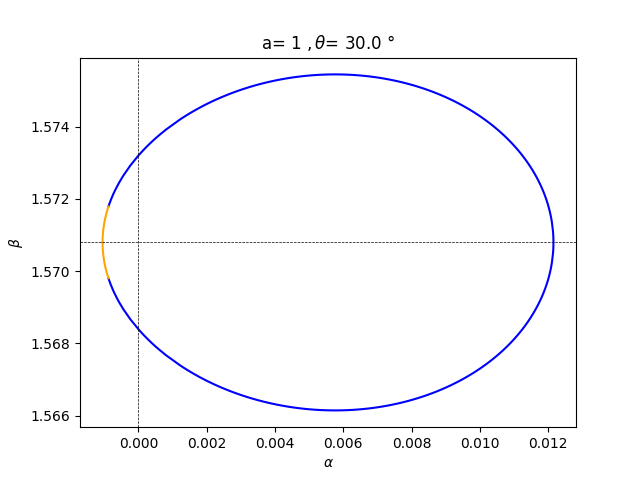}
   \includegraphics[width=4 cm, height=3.5 cm]{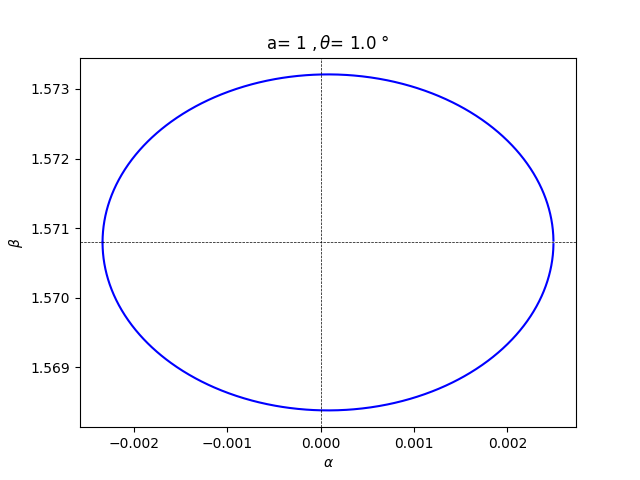}
   
  \caption{Critical angles in $\alpha-\beta$ plane in the extremal Kerr black hole ($a=1$). The red, green, blue and orange lines show ($\alpha_{1 (a)}, \beta_{1(a)}$), ($\alpha_{1 (b)}, \beta_{1(b)}$), ($\alpha_{1 (c)}, \beta_{1(c)}$) and ($\alpha_{2}, \beta_{2}$), respectively. We set $\mu=10^{-3}$ and we change $\theta = \pi/2, \pi/3, \pi/4, \pi/6, 0.017$. Let us stress that the seemingly sharp transitions in the first three panels are actually smooth as the curves in the remaining panels, although zooming in is necessary in order to appreciate it.}
\label{fig: fig3}
\end{figure}

\begin{figure}[!htbp]
  \centering
  \includegraphics[width=4 cm, height=3.5 cm]{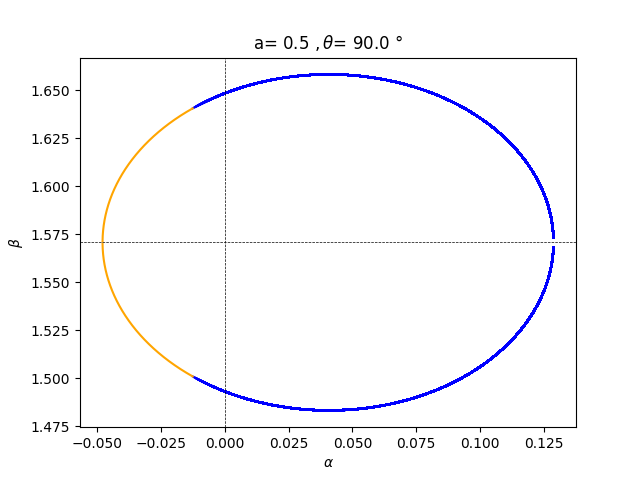}
  \includegraphics[width=4 cm, height=3.5 cm]{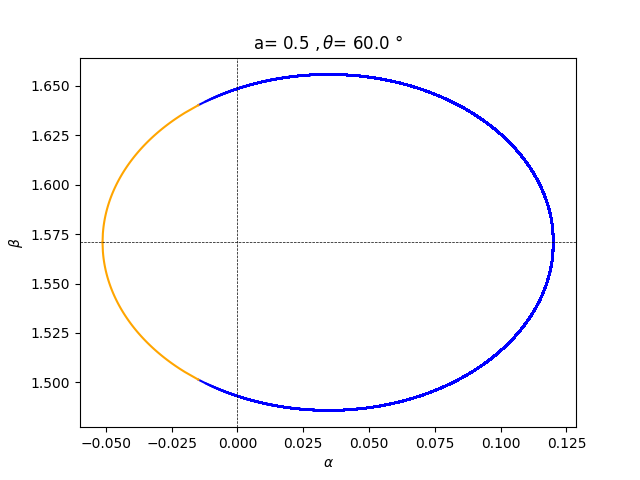}
  \includegraphics[width=4 cm, height=3.5 cm]{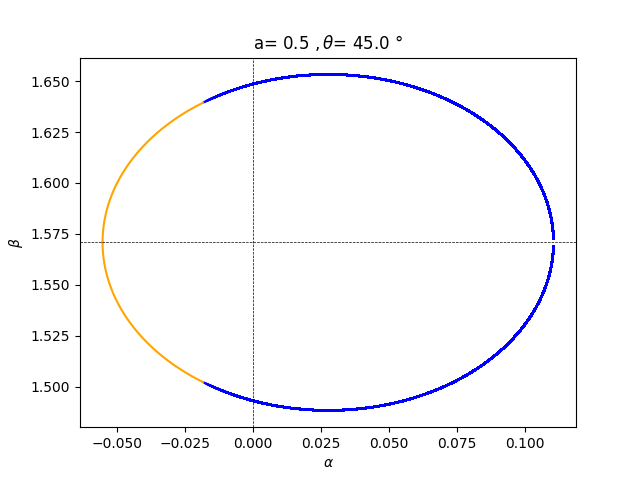}
   \includegraphics[width=4 cm, height=3.5 cm]{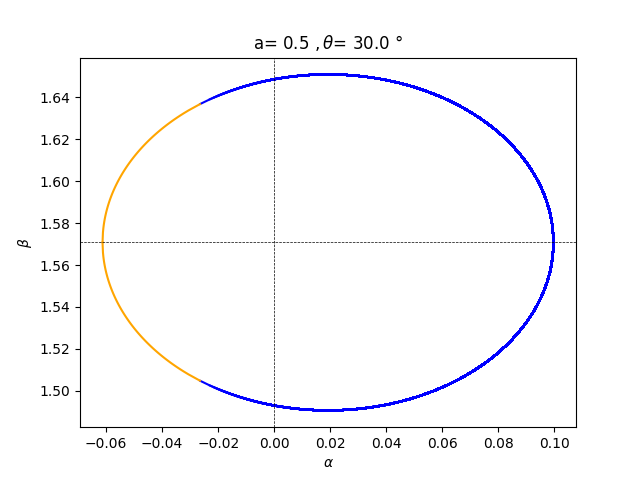}
   \includegraphics[width=4 cm, height=3.5 cm]{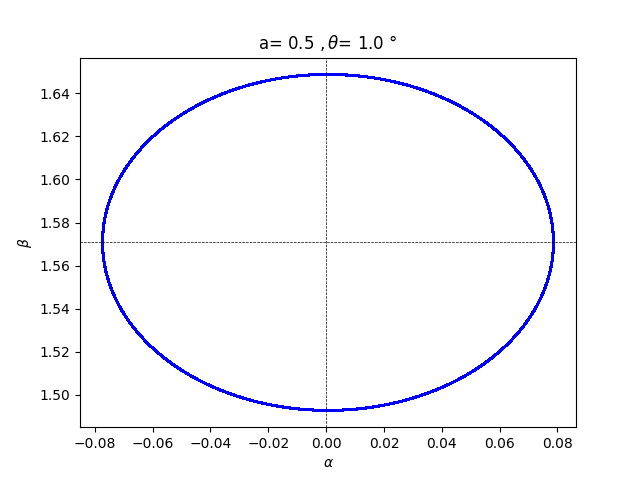}
   
  \caption{Critical angles in $\alpha-\beta$ plane in the Kerr black hole ($a=0.5$). The green, blue and orange lines show ($\alpha_{1 (b)}, \beta_{1(b)}$), ($\alpha_{1 (c)}, \beta_{1(c)}$) and ($\alpha_{2}, \beta_{2}$), respectively. We set $\mu=10^{-3}$ and we change $\theta = \pi/2, \pi/3, \pi/4, \pi/6, 0.017$.}
\label{fig: fig4}
\end{figure}

\begin{figure}[!htbp]
  \centering
  \includegraphics[width=4 cm, height=3.5 cm]{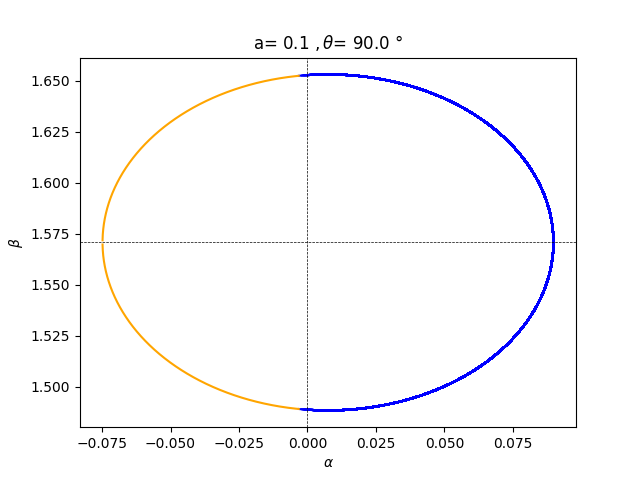}
  \includegraphics[width=4 cm, height=3.5 cm]{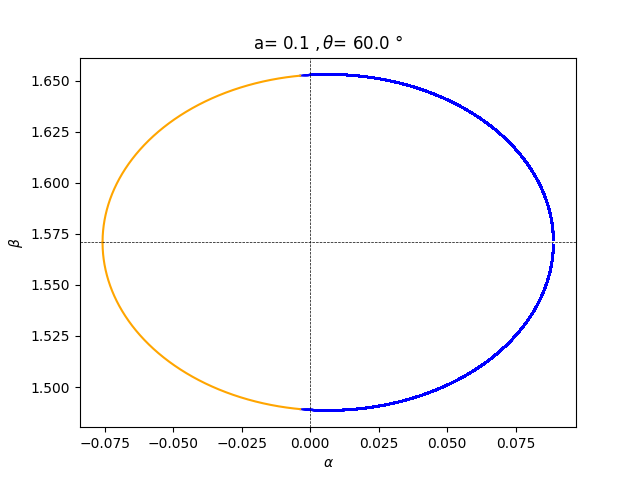}
  \includegraphics[width=4 cm, height=3.5 cm]{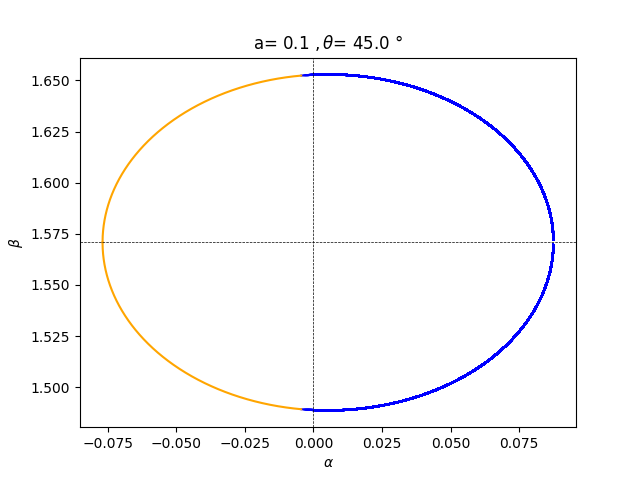}
   \includegraphics[width=4 cm, height=3.5 cm]{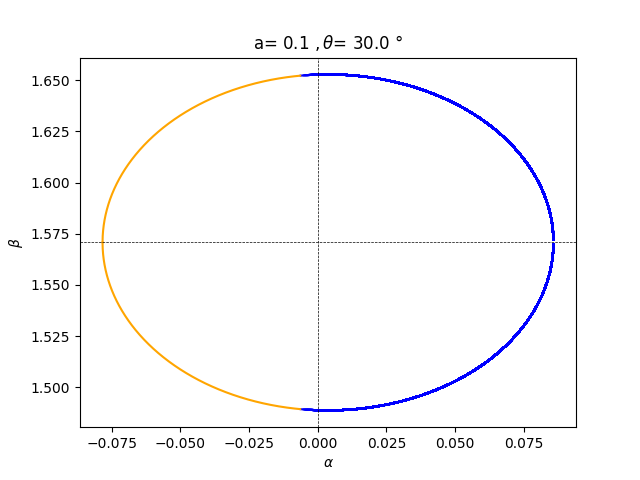}
   \includegraphics[width=4 cm, height=3.5 cm]{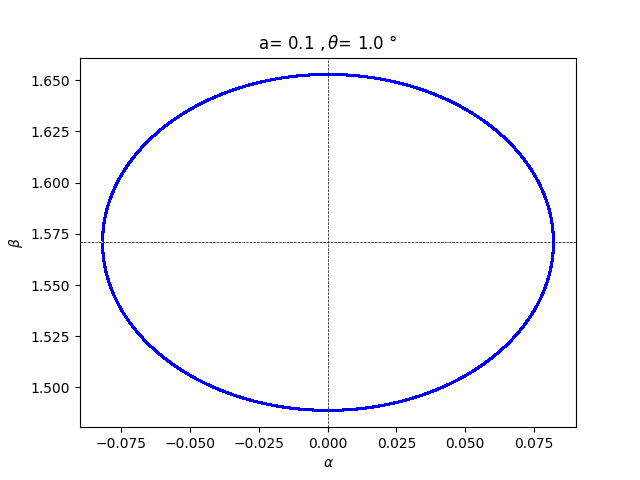}
   
  \caption{Critical angles in $\alpha-\beta$ plane in the Kerr black hole ($a=0.1$). The green, blue and orange lines show ($\alpha_{1 (b)}, \beta_{1(b)}$), ($\alpha_{1 (c)}, \beta_{1(c)}$) and ($\alpha_{2}, \beta_{2}$), respectively. We set $\mu=10^{-3}$ and we change $\theta = \pi/2, \pi/3, \pi/4, \pi/6, 0.017$.}
\label{fig: fig5}
\end{figure}

It is useful to look at numerical plots of the escape angles in the $\alpha-\beta$ plane, see Figs. \ref{fig: fig1} - \ref{fig: fig5}. First of all, we can plot the critical angles in the extremal Kerr black hole in order to check that we recover the main results of \cite{Ogasawara2019}. Then, we can go a step beyond and generalize the results of \cite{Ogasawara2019} for non-extremal rotation, showing in particular that in the case of very slow rotation the critical angles become symmetric as we expect in the case of spherical symmetry. It is also interesting to notice that for very small $a$ the critical angles do not vary much when $\theta$ varies, so the case of slow rotation behaves similarly to the case of spherical symmetry for which the emission probability does not depend on the plane of emission. 
%
\subsection{Escape probability}

\begin{figure}[!htbp]
  \centering
  \includegraphics[width=6cm, height=4.5 cm]{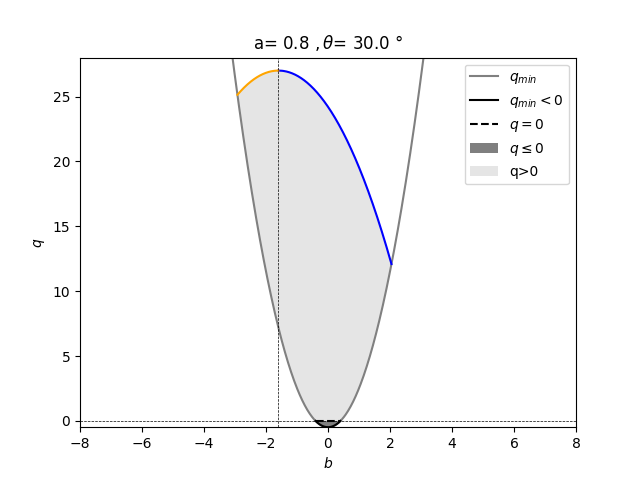}
  \includegraphics[width=6 cm, height=4.5 cm]{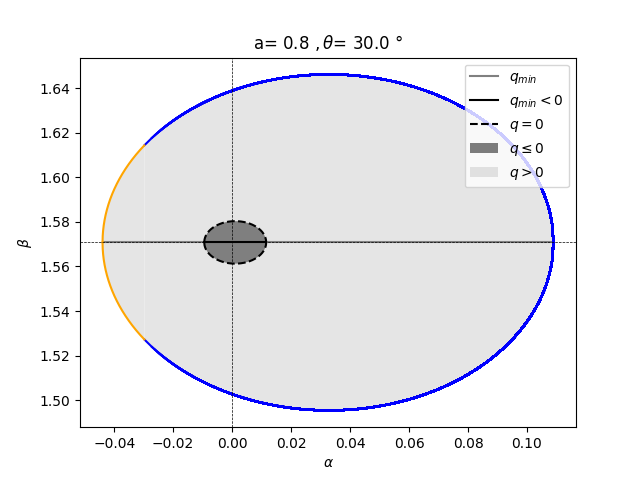}
  \includegraphics[width=6 cm, height=4.5 cm]{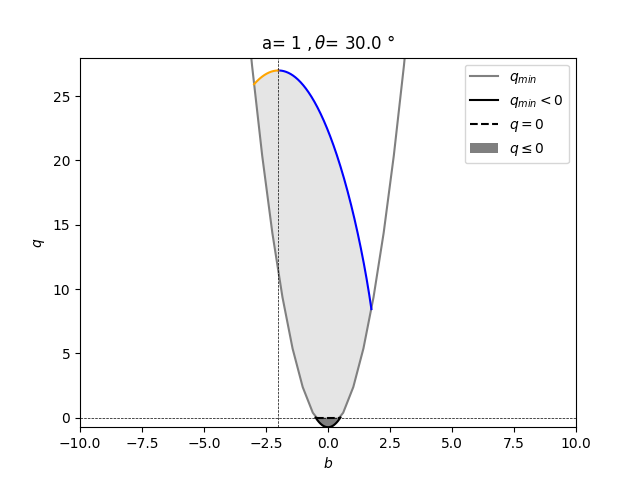}
   \includegraphics[width=6 cm, height=4.5 cm]{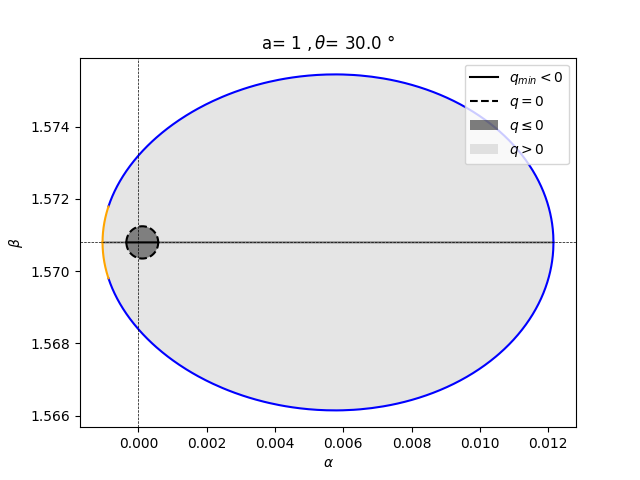}
   
  \caption{Illustration of the change of variables from $(b,q)$ to $(\alpha,\beta)$ used for the calculation of the escape probability, for both non-extremal ($a=0.8$) and extremal ($a=1$) situations. The region in which trajectories have $q<0$ is highlighted in each of these figures.}
\label{fig: qneg}
\end{figure}

We can now evaluate the escape probability $P$ for a photon in Eq. \eqref{PP}.  Let $\alpha_{\rm min}$ and $\alpha_{\rm max}$ be the minimum and the maximum of the critical angle $\alpha$, and $\beta_{\rm min}(\alpha)$ and $\beta_{\rm max}(\alpha)$ be the minimum and the maximum of the critical angle $\beta$ for each given $\alpha$. Assuming that $S$ is convex, we obtain 
\begin{equation}
P= \frac{1}{4 \pi} \int_{\alpha_{\rm min}}^{\alpha_{\rm max}} \text{d}\alpha \int_{\beta_{\rm min}}^{\beta_{\rm max}}\text{d}\beta \,\sin\beta = \frac{1}{2 \pi} \int_{\alpha_{\rm min}}^{\alpha_{\rm max}} \text{d}\alpha \,\cos\beta_{\rm min}(\alpha),
\label{prob_def}
\end{equation}
where we have used $\beta_{\rm max}=\pi-\beta_{\rm min}$. In other words, we are performing a change of variables from $(b,q)$ to the angles $(\alpha,\beta)$ through Eq. \eqref{eq:scab}, and then integrating in the latter variables. Fig. \ref{fig: qneg} below provides a graphical representation of this change of variables, together with the different regions inside the escape cones with $q>0$ and $q<0$. This change of variables is a two-dimensional mapping in which $q=q_{\rm min}$, being defined by the condition $\hat{\Theta}=0$, is mapped to a line of constant angle $\beta=\pi/2$ as it can be checked using Eq. \eqref{eq:scab}. It follows then that the integration in the $(\alpha,\beta)$ variables is exhaustive, in the sense that all the escaping trajectories are taken into account, regardless of the sign of $q$.

Using the critical angles ($\alpha_i,\beta_i$) and the relation between $q$ and $r_i$ given in Eq. \eqref{eq13}, we can change the integration variable $\alpha$ to $r_i$ as 
\begin{equation}
P=P_1+P_2,
\end{equation}
where
\begin{equation}
P_i=\frac{(-1)^i}{2 \pi} \int_{q_{i,{\rm min}}}^{q_{\rm max}} \text{d}q \frac{\text{d}\alpha_i}{\text{d}q} \cos\beta_i=\frac{(-1)^i}{ 2 \pi} \int_{r_{i,{\rm min}}}^{r_{\rm max}} \text{d}r_i \frac{\text{d}\alpha_i}{\text{d}r_i} \cos\beta_i.
\end{equation}
In particular, using the definition of the critical angles we can reduce $P_1$, if $q_{1,{\rm min}}<f(r_+)$,  as
\begin{equation}
P_1=
\begin{cases}
\!\begin{aligned}
-\frac{1}{2 \pi} \int_{r_{1,{\rm min}}}^{r_+} \text{d}r_1 \frac{\text{d}\alpha_{1 (a)}}{\text{d}r_1} \cos\beta_{1 (a)}-&\frac{1}{2 \pi} \int_{r_+}^{r_\star} \text{d}r_1 \frac{\text{d}\alpha_{1 (b)}}{\text{d}r_1} \cos\beta_{1 (b)}\\
& -\frac{1}{2 \pi} \int_{r_\star}^{r_{\rm max}} \text{d}r_1 \frac{\text{d}\alpha_{1 (c)}}{\text{d}r_1} \cos\beta_{1 (c)} 
\end{aligned} & \text{ (for $a=1$)}\\
\displaystyle-\frac{1}{2 \pi} \int_{r_{1,{\rm min}}}^{r_\star} \text{d}r_1 \frac{\text{d}\alpha_{1 (b)}}{\text{d}r_1} \cos\beta_{1 (b)}-\frac{1}{2 \pi} \int_{r_\star}^{r_{\rm max}} \text{d}r_1 \frac{\text{d}\alpha_{1 (c)}}{\text{d}r_1} \cos\beta_{1 (c)}  & \text{(for $0<a<1$)}
\end{cases}
\end{equation}
and, if $q_{1,{\rm min}}>f(r_+)$ as
\begin{equation}
P_1=
\begin{cases}
\!\begin{aligned}
-\frac{1}{2 \pi} \int_{r_{1,{\rm min}}}^{r_\star} \text{d}r_1 \frac{\text{d}\alpha_{1 (b)}}{\text{d}r_1} \cos\beta_{1 (b)} -\frac{1}{2 \pi} \int_{r_\star}^{r_{\rm max}} \text{d}r_1 \frac{\text{d}\alpha_{1 (c)}}{\text{d}r_1} \cos\beta_{1 (c)} 
\end{aligned} & \text{ (for $a=1$)}\\
\displaystyle-\frac{1}{2 \pi} \int_{r_{1,{\rm min}}}^{r_\star} \text{d}r_1 \frac{\text{d}\alpha_{1 (b)}}{\text{d}r_1} \cos\beta_{1 (b)}-\frac{1}{2 \pi} \int_{r_\star}^{r_{\rm max}} \text{d}r_1 \frac{\text{d}\alpha_{1 (c)}}{\text{d}r_1} \cos\beta_{1 (c)}  & \text{(for $0<a<1$)}
\end{cases}
\end{equation}
 where
\begin{equation}
\frac{\text{d}\alpha_{1 (a)}}{\text{d}r_1}=- \frac{\sigma_r \sin\theta}{2b(r_+) \Sigma \sqrt{\Delta}} \left[ \left(\frac{1}{\sqrt{\hat{R}A}}-\frac{\sqrt{\hat{R}A}sin^2\theta}{(\hat{R}A \sin^2\theta+b^2(r_+)\Sigma^2\Delta)} \right)A \frac{\text{d}\hat{R}}{\text{d}r_1}\right]
\end{equation}
and it is useful to note that some of the integrands in the above integrals have a common form
\begin{equation}
 \begin{split}
h(x)=& - \frac{\sigma_r \sin\theta}{2b(r_1) \Sigma \sqrt{\Delta}} \Biggl[ \left(\frac{1}{\sqrt{\hat{R}A}}-\frac{\sqrt{\hat{R}A}\sin^2\theta}{(\hat{R}A \sin^2\theta+{b(r_1)}^2\Sigma^2\Delta)}\right)A \frac{d\hat{R}}{dr_1}\\ &\quad- 2 b(r_1) \frac{db(r_1)}{dr_1} \frac{\Sigma^2 \Delta \sqrt{\hat{R}A}}{(\hat{R}A \sin^2\theta+{b(r_1)}^2\Sigma^2\Delta)} \Biggr]
\end{split}
\end{equation}
where 
\begin{equation}
h(x)= \frac{d\alpha_{1 (b)}}{dr_1} \big\vert_{r_1=x} = \frac{d\alpha_{1 (c)}}{dr_1} \big\vert_{r_1=x}= \frac{d\alpha_{2}}{dr_2} \big\vert_{r_2=x}
\end{equation}
and 
\begin{equation}
g(x)= h(x) \cos\beta_{1 (b)} = h(x) \cos\beta_{1 (c)} = h(x) \cos\beta_{2}
\end{equation}
Finally from eq. \eqref{prob_def}, if $q_{1,{\rm min}}<f(r_+)$, $P$ is given by 
\begin{equation}
P=
\begin{cases}
\displaystyle-\frac{1}{2 \pi} \int_{r_{1,{\rm min}}}^{r_+} \text{d}r_1 \frac{d\alpha_{1 (a)}}{\text{d}r_1} \cos\beta_{1 (a)}-\frac{1}{2 \pi} \int_{r_+}^{r_{2,{\rm min}}} \text{d}x\, g(x) & \text{(for $a=1$)}\\
\displaystyle-\frac{1}{2 \pi} \int_{r_{1,{\rm min}}}^{r_{2,{\rm min}}} \text{d}x\, g(x) & \text{(for $0<a<1$)}
\end{cases}
\label{P}
\end{equation}
while if $q_{1,{\rm min}}>f(r_+)$ is given by
\begin{equation}
\begin{array}{ll}
\displaystyle P=-\frac{1}{2 \pi} \int_{r_{1,{\rm min}}}^{r_{2,{\rm min}}} \text{d}x\, g(x) & \text{(for $ 0<a <1$ and $a=1$)}
\end{array}
\label{Pi}
\end{equation}

\begin{figure}[!htbp]
  \centering
  \includegraphics[width=5 cm, height=3.75 cm]{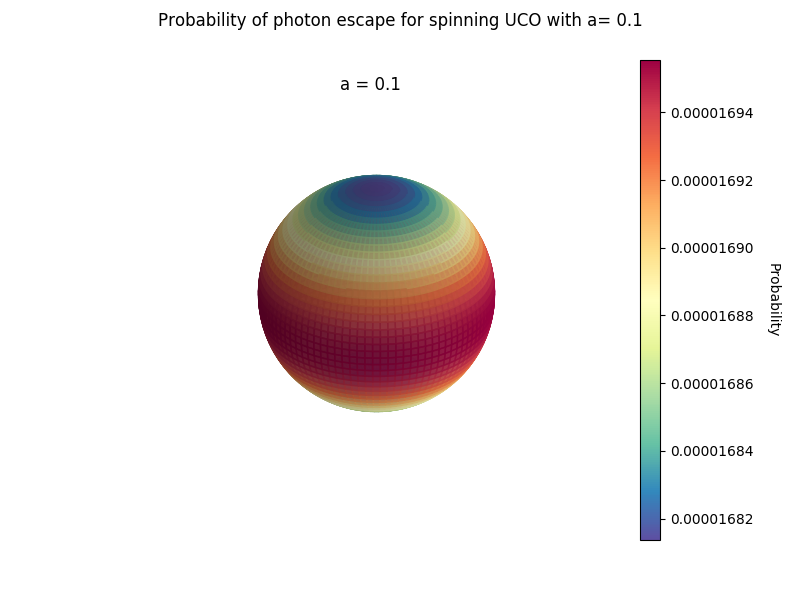} 
  \includegraphics[width=5 cm, height=3.75 cm]{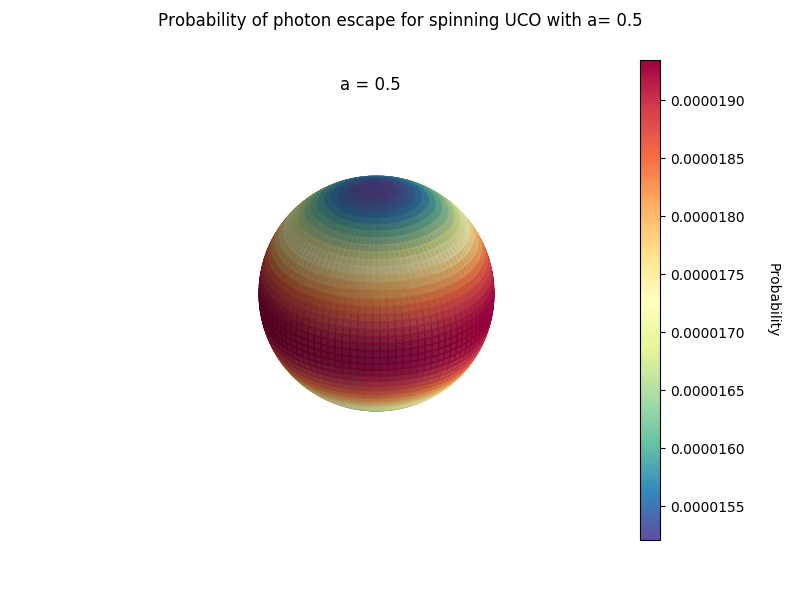} 
  \includegraphics[width=5 cm, height=3.75 cm]{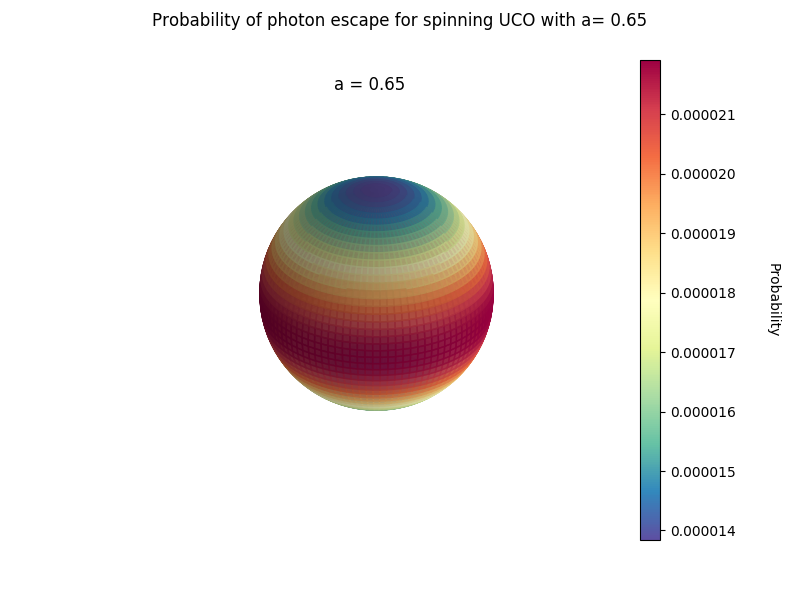} 
  \includegraphics[width=5 cm, height=3.75 cm]{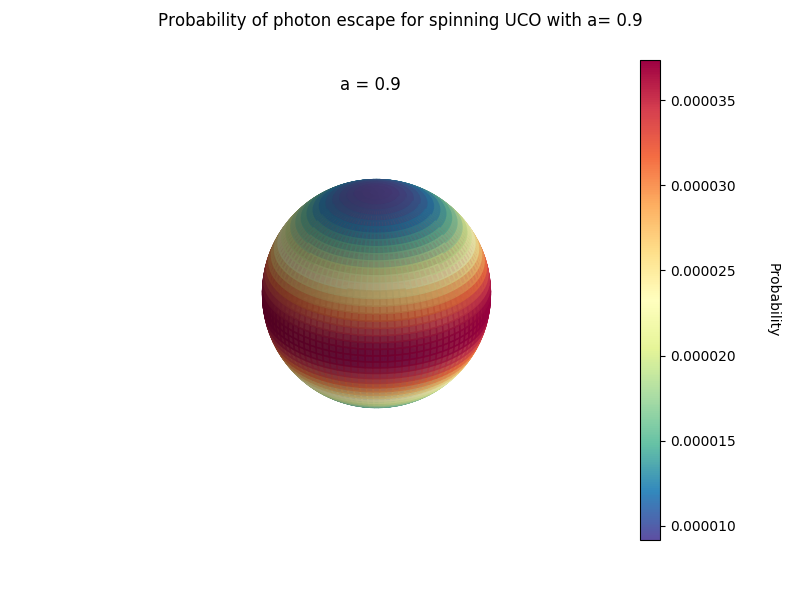} 
  \includegraphics[width=5 cm, height=3.75 cm]{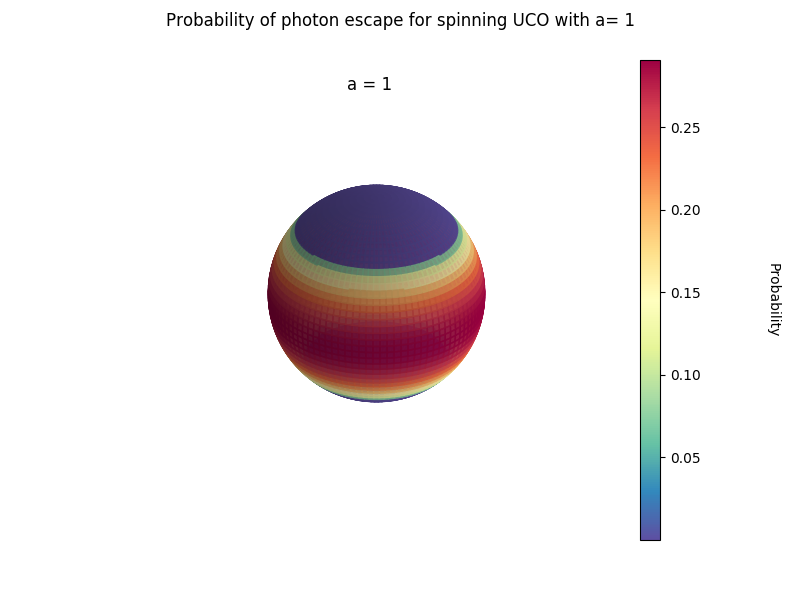} 
  \includegraphics[width=5 cm, height=3.75 cm]{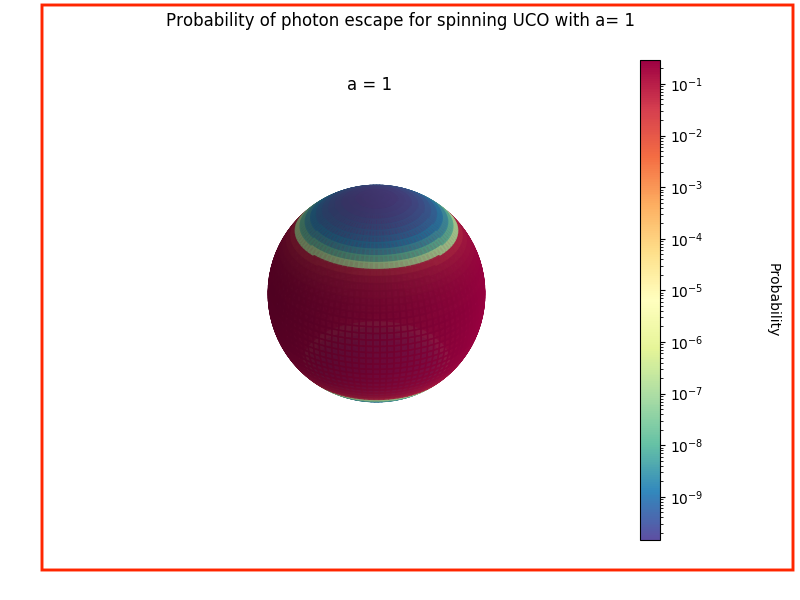} 
  
  \caption{Visualizations of photon escape probability for different values of $a$. Each panel has its own color bar, different from the others. In particular the case of $a=1$ (central and right bottom panels) is presented both in normal and logarithmic color scale. The latter clearly shows that even in the extremal case there is a region with the same value of probability of the non-rotating case and, moreover,  there is a region where the escape probability becomes smaller than this value.}
  \label{fig:P1a}
\end{figure}

\begin{figure}[!htbp]
  \centering
  \includegraphics[width=15 cm, height=9 cm]{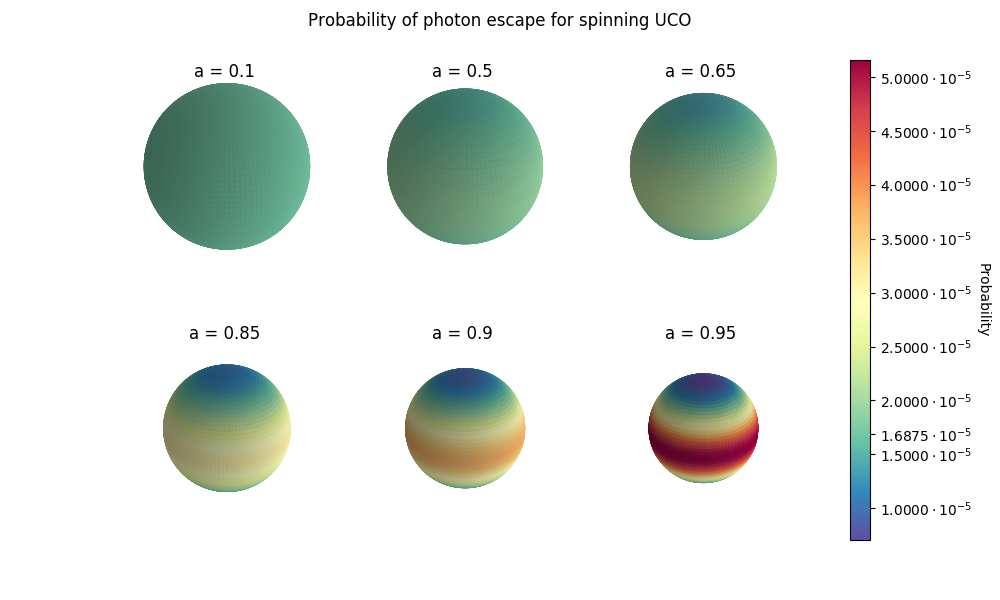}

  \caption{Visualizations of photon escape probability for different values of $a$, normalized to the same color scale. The value $P=1.6875 \cdot 10^{-5}$ corresponds to the case $a=0$.}
  \label{fig:P2a}
\end{figure}

The expressions above can be evaluated numerically for different values of $a$ and $\theta$, with the numerical results being showed in \cref{Tables}.
Moreover, in Figs. \ref{fig:P1a} - \ref{fig:P2a} we present some visualizations of the obtained results. Color maps allow us to better visualize the numerical data on the sphere that corresponds to the representation of the $r_\star$ surface in Boyer-Lindquist coordinates.
All the visualizations are represented for the case of $\mu=10^{-5}$ and present a color bar that allows us to interpret the color scale on the base of numerical results.


It is interesting to note that, if we take the (homogeneous) value of the non-rotating escape probability as reference, the value of the escape probability becomes smaller around the poles when $a\neq0$. This may seem counterintuitive, given that one might expect that the escape probability at the poles should have the same value as in the non-rotating case. However, one must take into account that the escape probability is not a property that is defined along a single line, but rather is a property of the set of light rays that are emitted isotropically at a given point (in other words, it is a property defined on a two-dimensional submanifold). Even if the point of the emission is located on the axis of rotation, most of the emitted rays will diverge from the rotation axis. The latter rays experience the effect of angular momentum which, as we have calculated, leads to the existence of turning points for a fraction of these rays. This makes the escape cone narrower, thus explaining why the escape probability around the pole is smaller than in the non-rotating case. 

\subsection{Integration on total surface} \label{sec:total}

Until now we have obtained values of probability which depend on the point of emission. Namely, we have chosen a point on a surface at a fixed value of $r=r_\star$ and $\theta$. In this section we consider the integrated emission from the surface at $r=r_\star$. In order to do this we just have to integrate numerically in $\text{d}\theta$ all the values of probability that we have obtained above. The algorithm that we apply is as follows:

\begin{enumerate}
\item We divide the angle range $[\ang{1},\ang{90}]$ in $n$ intervals of length $\Delta\theta=(\pi/2-0.017)/n$.
\item Note that we are not taking into account the angles in the interval $[\ang{0},\ang{1}]$. On the pole ($\theta=0$), the Kerr metric has a coordinate singularity and we cannot integrate total probability from the pole to the equator in the coordinate system that we are considering; for this reason we start integrating from $\theta=\ang{1}$. In this way we introduce uncertainties due to the fact that we are neglecting the contribution from angles in the range $[\ang{0},\ang{1}]$. Assuming that the probability presents a monotonous behavior with the angle, which is verified in the interval $[\ang{1},\ang{90}]$, an upper bound to this error is given by $(\theta_1-\theta_0)\abs{\sin\theta \ \Sigma (\theta) \ P(\theta)}_{\theta=\theta_\star}$, where $\theta_\star$ corresponds to the maximum of this expression.
\item We evaluate a table of values, like the ones presented in the previous section, but for a certain value of $a$ and $r=r_\star$ fixed and for each value of $\theta$ in this range. These values are multiplied by the corresponding surface element $dS$ in Eq. \eqref{surf} and by $\Delta \theta$.
\item We calculate $\sum _{\theta} P(\theta) dS(\theta) \Delta\theta$ and divide it by the total surface of these ellipsoids $S$ (given by Eq. \eqref{S_K}) to obtain the final value of total integrated probability  on the surface:
\begin{equation}
P_{\rm tot} = \frac{\sum _{\theta} P(\theta) dS(\theta) \Delta\theta}{S}
\label{P_tot}
\end{equation}
 \end{enumerate}
The values of this probability obtained for different values of $a$, for $\mu= 10^{-5}$ and for $n=1000$ are reported in Tab. \ref{tab: AA5}. We also calculated this probability for different values of $\mu$, namely $\mu=10^{-3}$ and $\mu=0.1$, in order to see if the integrated probability behaves in the same way also for different values of compactness. These results show that in slow rotating case the integrated probability is compatible with the zero rotation limit, showing that, as the rotation decreases, the probability tends to be more and more independent from the emission point on the surface.
However, as the spin parameter increases, the integrated probability is dominated by the contribution coming from the region around the equator, which is larger than in the Schwarzschild case. Our analysis shows that such an increase dominates over the decrease of the probability in the polar regions.
Until now, we have calculated the integrated probability also for the case of extremal black hole, the maximally rotating black hole, for which $a=1$, albeit the aforementioned Thorne limit makes this case more of theoretical interest than of astrophysical relevance.

Before ending this section let us provide a physical interpretation of the numerical values obtained. It is well known that in the case of a rotating Kerr black hole, there are two unstable circular orbits that could exist in the equatorial plane. As we have just understood, the behavior of the integrated probability in the case of very fast rotation is dominated by the equatorial contribution so we can focus on what happens to these unstable orbits. Although such orbits are unstable, they are nevertheless important from a physical point of view because they define the boundary between capture and non-capture of a cross-section of light rays by the Kerr black hole
 \cite{Pugliese2011}.
One of this circular orbits is a prograde orbit moving in the same direction as the black hole's rotation, while the other is a retrograde orbit moving against the black hole's rotation. Their radii are, respectively, given by \cite{Teo2003}:
\begin{equation}
\begin{split}
&r_1= 2 \left[1+ \cos\left(\frac{2}{3} \arccos(-\abs{a})\right)\right],\\
&r_2= 2 \left[1+ \cos\left(\frac{2}{3} \arccos(\abs{a})\right)\right].\\
\end{split}
\end{equation}
These are sometimes referred to as the Kerr geometry light rings , which can in principle touch the surface of the ultracompact object, so that photons departing from the surface can be very near to the tip of the potential. Assuming that to a good approximation the geometry outside the object is given by Kerr geometry, we can study for which value of $a$ the internal light ring touches the surface of the object, that corresponds to  $r_\star=r_+(1+\mu)$, with a given $\mu \ll 1$.

In other words, we search the value of $a$ for which the co-rotating light ring touches the surface, defined fixing a certain value of $\mu$, that means we are interested in calculating the value of $a$ for which $r_1=r_\star$.
The equation 
\begin{equation}
f_1(a)=\frac{(1+\sqrt{1-a^2})(1+\mu)}{2}-1=\cos\left(\frac{2}{3} \arccos(-\abs{a})\right)=f_2(a)
\end{equation}
can be solved using bisection method, finding the values of $a$ for which the co-rotating circular orbit touches the surface (see Tab. \ref{tab:valuea}).

\begin{figure}[!htbp]
  \centering
  \includegraphics[width=5 cm, height=4 cm]{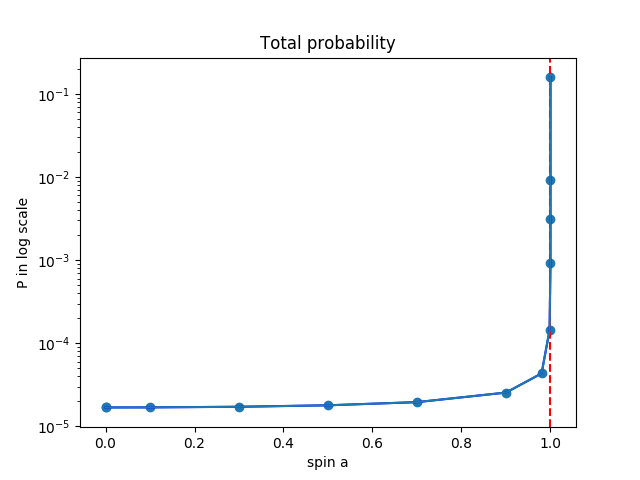}
  \includegraphics[width=5 cm, height=4 cm]{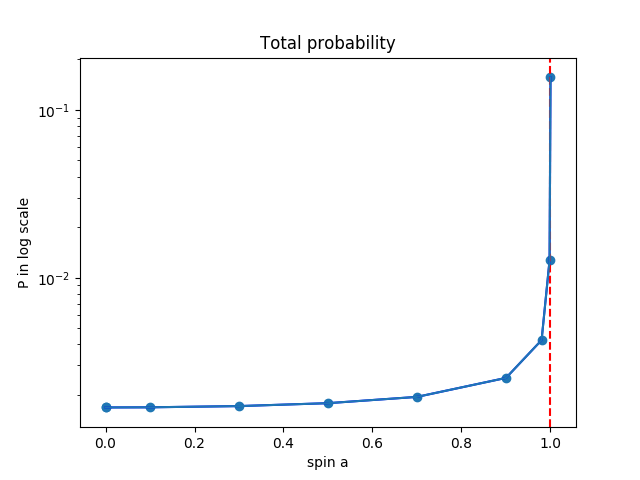}
  \includegraphics[width=5 cm, height=4 cm]{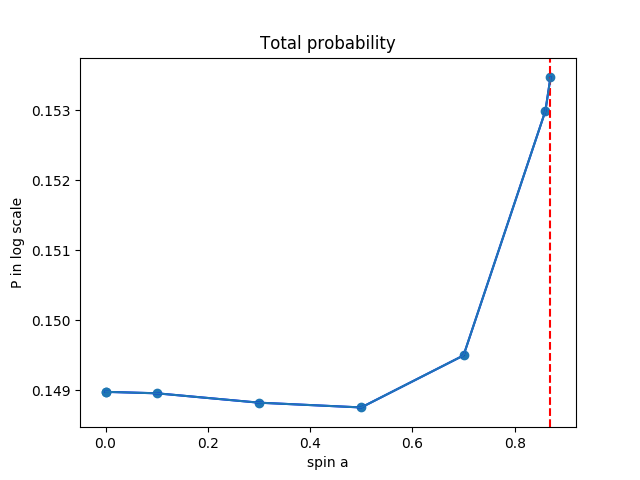}
  
  \caption{Total Probability of photon escape from surfaces at fixed $r=r_\star$ for $\mu=10^{-5}, \ 10^{-3}, \ 10^{-1}$ respectively. Error bars are not visible because the uncertainties are very small.}
  \label{fig: P1}
\end{figure}

 \begin{table}[h]
\begin{tabular}{@{\qquad}c@{\qquad} | @{\qquad}c@{\qquad}}
\hline
\hline
$\mu$ & $a$\\
\colrule
$10^{-5}$ & $0.9999999979118$\\
$10^{-3}$ & $0.9999792249829$\\
$10^{-1}$ & $0.8697788466062$\\
\hline
\hline
\end{tabular}
\label{tab:valuea}
\caption{Values of $a$ for which the co-rotating circular orbit touches the surface}
\end{table}
These values corresponds to the values of $a$ for which probability starts to decrease dramatically as we can see in Fig. \ref{fig: P1} (red dashed line corresponds to the value of $a$ for which co-rotating light ring touches the surface).
 %

 \section{Sagittarius A*}  \label{sec:SgrA*}
 
 Let us study the particular case of Sagittarius A* (Sgr A*), the radio point-source associated with the dark mass located at the center of the Milky Way. Near-infrared (NIR) observations of massive stars in its vicinity have provided direct mass and distance measurements, $M=(4.5 \pm 0.4) \times 10^6 \ M_{\odot}$ and $D=8.4 \pm 0.4$ kpc \cite{Broderick2009}. With a luminosity of $10^{36} \ \text{erg} \ \text{s}^{-1}$, it is substantially under-luminous relative to its limiting Eddington luminosity:
 \begin{equation}
 L_{\rm Edd}= 3.3 \times 10^4 \frac{M}{M_{\odot}} L_{\odot} \simeq 6 \times 10^{44} \ \text{erg} \  \text{s}^{-1}.
 \label{Ledd}
  \end{equation}
Providing measures about the spin of the nearest supermassive black hole is more complicated. While mass measurements can be made at a large distance from an object, spin requires a probe which is close in, namely within $10 \ M$. The standard approach is to identify the inner edge of the accretion disk with the ISCO and then convert that radius to spin.
Another way to measure the mass and spin of the black hole is Quasi-Periodic Oscillations of the hot plasma spots or clumps orbiting an accreting black hole that contain information on these parameters. Their interpretation reveals the Kerr metric rotation parameter $a=0.65 \pm 0.05$ for Sgr A* \cite{Dokuchaev2014}.

As briefly reviewed in Sec. \ref{sec:back}, if the two assumptions of thermalization and steady state hold the emission of the ultracompact object can be calculated: it is given by a thermal distribution with a temperature determined by the accretion rate $\dot{M}$. It has been shown that the emission of Sgr A* in the infrared is about $10^{-2}$ times this theoretical estimate. It was then concluded \cite{Narayan1997,Narayan2002,McClintock2004,Narayan2008,Broderick2009} and often quoted in literature that it is not possible that Sgr A* has a surface, and therefore that it must have an horizon (however, this conclusion has been substantially revised afterwards \cite{Abramowicz2002,Lu2017,Carballo-Rubio2018,Carballo-Rubio2018b,Cardoso2019}). In revising the above derivation an obvious starting point is to reconsider the two assumptions presented before. 
In the case of Sgr A*, the object is not rotating very fast so we can assume, in a first approximation, that the case of slow rotation behaves similarly to Schwarzschild case (Fig. \ref{fig:P2a} shows as the probability of photon escape looks pretty close to be uniform all over the emitting surface) and so we can assume thermalization.
As regards steady state assumption, even if black holes explicitly violate this condition because of the unradiated kinetic energy advected across the horizon and then added to the mass of the object, it is reasonable to expect that any black hole alternative will reach some sort of steady state if given enough time \cite{Broderick2009}; the main question is then how much time is needed for the assumption of steady state to be reasonable.

Let us now revise the calculation of the time at which steady state can be reached. The initial configuration is the same presented in \cite{Carballo-Rubio2018b}: the accretion disk starts pumping energy into the slow rotating ultracompact object, while the energy emission of the latter is considered negligible before accretion begins.
We shall introduce $\dot{E}$ as the amount of energy emitted per unit time by the ultracompact object measured at the location of the accretion disk $r=R_{\rm disk}$. We want to describe the evolution of the system and so we need to keep into account two effects. 
First of all, it is useful to evaluate the time until the first ingoing radial null geodesics\footnote{We assume that all propagating energy is carried along null geodesic.} can bounce back at the surface $r=r_\star$ and return to the accretion disk because until this moment the energy emitted $\dot{E}$ remains negligible. We assume that in the case of slow rotation this time is of the same order of magnitude of  $\mathcal{O}(10) \times r_+$, in analogy with Schwarzschild case. Hence, this timescale is essentially the light-crossing time of the ultracompact object.

Then, there is a second effect to keep into account. Outgoing null geodesic are strongly lensed, which implies that a fraction of them do not escape and fall again onto the surface of the ultracompact object. This effect is unavoidable due to the inherently inelastic nature of the process that is necessary for thermalization to take place: the energy falling from the accretion disk is absorbed by the ultracompact object in the first place, and then emitted. If the probability of photon escape is uniform all over the surface, like for Schwarzschild, then particles would fall onto the surface of the object and then would be reemitted uniformly and so the analysis of the process would not be affected, but if the object is rapidly rotating we should take into account this effect in a more rigorous way, following  the process step by step.  
In any case, for Sgr A* the remaining energy follows highly curved trajectories and is reabsorbed by the ultracompact object in a timescale that can be calculated numerically and, in analogy with Schwarzschild case, we can assume it is controlled by the horizon radius, being $\mathcal{O}(10) \times r_+$. Then a repetition of this process takes place, until eventually all the energy is radiated away.
 
In order to make the calculation tractable, we can consider discrete intervals with their size given by the characteristic timescale $\tau_s=\mathcal{O}(10) \times r_+$ starting a $t=T_{\rm bounce}$. During each of these intervals, the mass that the accretion disk is ejecting into the object is given by $\dot{M} \tau_s$. In the first interval after $T_{\rm bounce}$, the amount of outgoing energy that reaches the accretion disk is given by the corresponding fraction of the first injection of energy,
\begin{equation}
E_1=P(\mu) \dot{M} \tau_s
,
\label{E1}
\end{equation}
where $P(\mu)$ is the integrated probability obtained above.
 
During the second interval, one would get the same fraction of the energy corresponding to the second injection plus a fraction of the remaining energy from the first injection:
 \begin{equation}
  \begin{split}
 E_2&=\left[P(\mu)+ P(\mu)(1-P(\mu))\right] \dot{M} \tau_s \\
 &= E_1+(1-P(\mu))E_1
\end{split}
 \end{equation}
 In general, one can show that 
  \begin{equation}
 E_n=\sum_{k=1}^{n} \epsilon_k
 \label{serie}
 \end{equation}
 where the particle energies can be determined from the recurrence relation
  \begin{equation}
\epsilon_{k+1}=(1-P(\mu))\epsilon_k,\ k \geq 1
\label{relricorrenza}
 \end{equation}
 with the seed $\epsilon_1=E_1$ given in Eq. \eqref{E1}. Summing the geometric series, it follows then that 
   \begin{equation}
 E_n=P(\mu) \dot{M} \tau_s \sum_{k=0}^{n-1} \left(1-P(\mu)\right)^{k}= \dot{M} \tau_s \left[1-(1-P(\mu))^n\right]
 \end{equation}
The accretion rate $\dot{M}$ is obtained dividing the mass accreted in each of these intervals by $\tau_s$. Therefore, let us analogously define $\dot{E}_n=E_n/\tau_s$. When $\tau_s \ll T$, the timescale during which the accretion rate $\dot{M}$ is roughly constant, we can formally take the limit in which the size of the time intervals goes to zero and therefore $\dot{E}_n$ becomes a function of a continuous variable, $\dot{E}(t)$, which can be written in terms of the continuous variable $t \in [T_{\rm bounce}, T]$ as
  \begin{equation}
\frac{\dot{E}(t)}{\dot{M}}=1-\left[1-P(\mu)\right]^{(t-T_{\rm bounce})/\tau_s}
\label{order}
 \end{equation}
 In the limit $r_\star \rightarrow r_+$ ($\mu \rightarrow 0$) one has $\dot{E}/\dot{M} \rightarrow 0$. 
This limit illustrates that a relativistic lensing effect cannot be ignored for $\mu \ll 1$, and can indeed spoil the stabilization of the composite system into a steady state. In particular, for Sgr A* the typical timescale for the variation of its accretion rate is set by the Eddington timescale 
\begin{equation}
T=\frac{Mc^2}{L_{\rm Edd}}\simeq 3.8 \times 10^8 \ \text{yr}
\end{equation}
where $L_{\rm Edd}$ is given by Eq. \eqref{Ledd}.
Hence, given that the emission of Sgr A* is at most $10^{-2}$ times that predicted under steady state assumption, we can write 
\begin{equation}
\frac{\dot{E}}{\dot{M}}\Bigg|_{t=T} = 1-\Big[1-P(\mu)\Big]^{(T-T_{\rm bounce})/\tau_s}  \leq \mathcal{O}(10^{-2})
\end{equation}
We can evaluate this quantity for different values of $\mu$ and in this way we obtain that
\begin{equation}
\mu \leq \mathcal{O}(10^{-16}).
\label{eqmu1}
\end{equation}

\begin{figure}[!htbp]
  \centering
  \includegraphics[width=8 cm, height=6.5 cm]{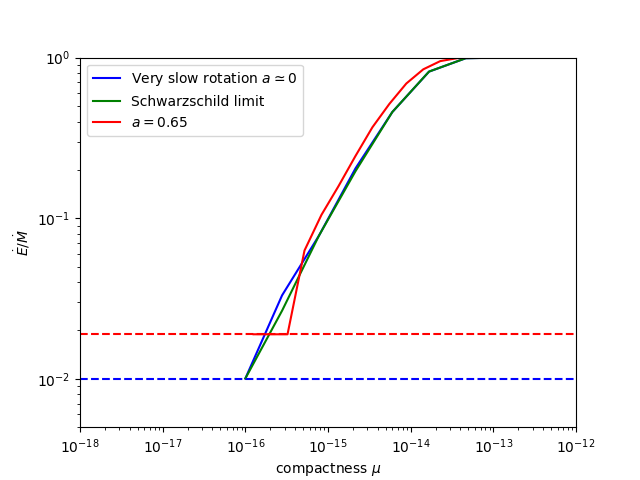}
  \includegraphics[width=8 cm, height=6.5 cm]{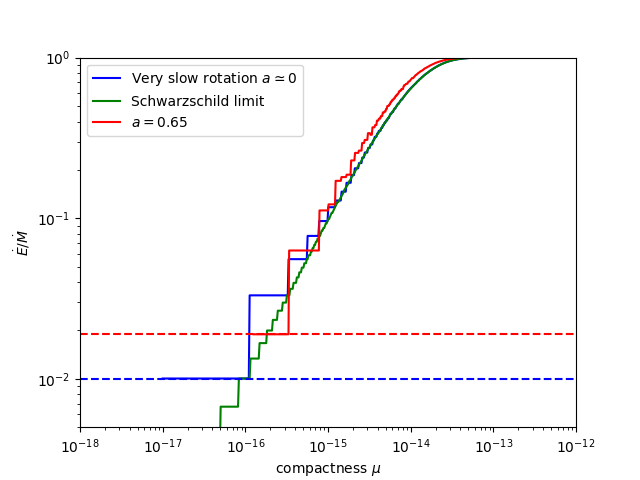}
  
  \caption{$\dot{E}/\dot{M}$ for different values of $\mu$ for Sgr A* ($a=0.65$), for a slow rotating object ($a=10^{-7}$) and for non rotating object ($a=0$). The last case is solvable analytically.\cite{Carballo-Rubio2018b} In left panel the values of $\mu$ are sampled with $n=20$ number of steps, while in the right panel with $n=500$.}
  \label{fig: mu}
\end{figure}

In Fig. \ref{fig: mu} we show the behavior of $\dot{E}/\dot{M}$ for different values of $\mu$ and we also plot the cases of $a \rightarrow 0$ (see \cite{Carballo-Rubio2018b}) and $a=10^{-7}$. Fig. \ref{fig: mu} (right panel) show the same plot where it has been increased the number of steps (the number of values of $\mu$) for which $\dot{E}/\dot{M}$ has been calculated. This allows us to check if its behavior is exactly the one of Fig. \ref{fig: mu} and so if the value for which the flux is $\mathcal{O} \simeq 10^{-2}$ then $\mu \leq \mathcal{O}(10^{-16})$. \\

\begin{figure}[!h]
  \centering
  \includegraphics[width=8 cm, height=6.5 cm]{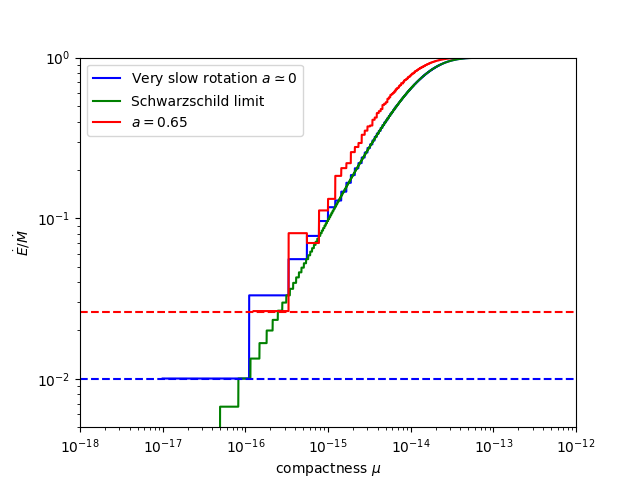}
  
 \caption{$\dot{E}/\dot{M}$ for different values of $\mu$ for Sgr A* ($a=0.65$), for a slow rotating object ($a=10^{-7}$) and for non rotating object ($a=0$), with probabilities estimated on the equatorial plane. The values of $\mu$ are sampled with $n=500$ number of steps.}
 \label{fig: mu3}
\end{figure}

In all this discussion, we have assumed that the probability of photons escaping from the ultracompact object has always the same value all over the surface and it corresponds to the integrated probability we have obtained in Sec. \ref{sec:total}.
However, in previous sections we have shown that probability changes with the angle respect to the pole and, in particular for the case of fast rotation, probabilities estimated on the pole and on the equatorial plane differ by about five orders of magnitude.
Let us now follow step by step all the calculations of this last section, in order to estimate the value of compactness for which $\dot{E}/\dot{M} \simeq \mathcal{O}(10^{-2})$, in the assumption that the probability on all the surface is equal to the probability on the equatorial plane. 
The results are shown in Fig. \ref{fig: mu3}. As we can see, this assumption does not change the results obtained above in a relevant way. The value for which the flux is $\mathcal{O} \simeq 10^{-2}$ is still of the same order of magnitude $\mu \leq \mathcal{O}(10^{-16})$. \\
Finally, we should compare results in Eq. \eqref{eqmu1} with theoretical values of $\mu$, obtained relating $\mu$ with the distance between the surface and the would-be horizon.
In spherical symmetry, for $\mu \ll 1$ and for a proper radial distance $\ell \ll r_s$ of the surface form $r_s$,  the relation is given by
\begin{equation}
\mu \simeq \frac{1}{4} \left(\frac{\ell}{r_s}\right)^2 \simeq 10^{-78} \left(\frac{M}{M_{\odot}}\right)^2 \left(\frac{\ell}{\ell_{\rm P}}\right)^2
\label{eqmu2}
\end{equation}
For Sgr A*, considering for instance $\ell \sim \ell_{\rm P}$, the value of $\mu$ is given by $\mu \sim 10^{-91}$.
So, Eq. \eqref{eqmu1} should be improved by about $75$ orders of magnitude in order to rule out these theoretical values on the basis of this argument alone (for complementary constraints that follow from a different argument, see \cite{Carballo-Rubio2018}). Moreover, we have not taken into account explicitly other phenomenological parameters that describe, for instance, absorption of electromagnetic waves. When these additional parameters are taken into account, the corresponding constraints become even weaker (in some cases, insignificantly weak) \cite{Carballo-Rubio2018b}.

It is interesting to compare our results with the recent EHT observations \cite{Akiyama2019}. Even if the EHT observations of Sgr A* are not available yet, the arguments made below apply equally to M87* (as well as any other supermassive black hole). Our main observation is that contrasting our estimate with these observations entails comparing our equations above for $\dot{E}/\dot{M}$ with the relative brightness of the central depression measured by the EHT, which for M87* has a relative value of $10^{-1}$ when compared with the bright disk in these images \cite{Akiyama2019}. Let us assume that the brightness of this disk is proportional to the accretion rate, namely
\begin{equation}
\alpha\dot{M}
\end{equation}
for some value of $\alpha$. This would imply that, for M87*, the EHT could only detect ratios $\dot{E}/\dot{M}$ that are greater than $10^{-1}\alpha$. Regardless of the final value of the ratios $\dot{E}/\dot{M}$ that can be detected by the EHT for a specific astrophysical system, the main message is that it will just represent an upper bound. Given that our equations above show that this ratio vanishes linearly with $\mu$ in the $\mu\rightarrow0$ limit, it follows that EHT observations cannot be used to discard sufficiently compact objects; in other words, no matter the sensitivity of the EHT, it is always possible to choose low enough values of $\mu$ that will ensure compatibility with these observations. Moreover, taking into more realistic models accounting for additional phenomenological parameters leads to even to bleaker scenarios \cite{Carballo-Rubio2018b}; for instance, even reasonably small absorptions coefficients are enough to strongly dampen the value of $\dot{E}/\dot{M}$.

Aside from these general observations, it is difficult to obtain more precise constraints. Taking into account the geometry of accretion disks, it is reasonable to assume that the fraction of photons that escape and are captured by the EHT is much greater than the fraction of photons accreted, namely $\alpha\gg1$. This would imply that current EHT observations cannot be used to extract meaningful constraints on $\mu$ whenever the latter is less than $\mathcal{O}(1)$. This would be a rephrasing of the statement that the EHT is mostly sensitive to physical processes taking place around the photon sphere, but cannot provide reliable information about processes taking place closer to the gravitational radius; see \cite{Carballo-Rubio2018b} for a more detailed discussion. In any case, this is certainly an interesting issue that deserves a separate and thorough analysis.

\section{Conclusions} \label{sec:con}

In this paper, we have presented a detailed analysis of the effect that rotation has on the geodesics around ultracompact horizonless objects. This has allowed us to extend the arguments originally presented in \cite{Narayan1997,Narayan2002,McClintock2004,Narayan2008,Broderick2009,Broderick2015}, constraining the properties of a hypothetical surface emitting electromagnetic radiation, to more realistic situations. The main ingredient of our analysis is the evaluation of the fraction of photons that escape to infinity when emitted isotropically from a given point on the surface as a function of the azimuthal angle with respect to the rotation axis of symmetry. The main features of this escape probability are:
\begin{itemize}
\item{The escape probability becomes increasingly anisotropic, as a function of the azimuthal angle, as the angular momentum increases. However, one has to reach relatively high values of the angular momentum for these anisotropies to become large (as defined below).}
\item{The escape probability decreases towards the axis of rotation and reaches its maximal value on the equatorial plane, with the maximal value being a monotonically increasing function with $a$. For the equatorial value to increase at least by a factor of 2 one needs $a\gtrsim 0.9$. On the other hand, the escape probability dips below its non-rotating value in an extended region around the poles.}
\end{itemize}

These features have the following implications for the conclusions that can be drawn for specific astronomical sources:
\begin{itemize}
\item{For values of the angular momentum below $a\simeq 0.9$, the emission from the surface is nearly isotropic, even in the presence of an anisotropic accretion disk. This follows from the redistribution of light rays after numerous cycles of emission and reabsorption after being lensed back to the surface, which washes away the initial angular dependence with which these light rays were injected from the accretion disk.}
\item{For higher values of angular momentum, the anisotropy in the emission from the surface and, in particular, the fact that it becomes smaller around the poles, makes necessary to consider the inclination of the source is needed in order to obtain reliable constraints; depending on the value of the inclination, the corresponding constraints on the properties of a hypothetical surface could become irrelevant. Moreover, these anisotropies may have an important effect on the onset of a steady state, which may even be disrupted or delayed. Hence, additional analyses are needed in order to understand whether the steady state assumption remains reasonable in these situations.}
\end{itemize}

It is interesting to point out that the two sources to which the arguments of \cite{Narayan1997,Narayan2002,McClintock2004,Narayan2008,Broderick2009,Broderick2015} have been applied, namely Sgr A* \cite{Broderick2009} and M87* \cite{Broderick2015}, present a very different behaviors. For Sgr A*, estimations of the spin yield results around between $a\simeq 0.4$ and $a\simeq 0.7$ \cite{Kato2010,Dokuchaev2014,Chashkina2014,Psaltis2015,Johannsen2015}, which makes this source belong to the first category in the itemization above. We have thus obtained reliable constraints on the radius of a hypothetical surface, which are nevertheless not strong enough to discard completely the existence of a surface (in fact, depending on the phenomenological parameters considered explicitly, these constraints could become insignificantly weak). On the other hand, recent estimations of the spin of M87* \cite{Tamburini2019} point to a much higher value $a\simeq 0.9$ and an inclination $i \simeq \ang{17}$ which implies that additional analyses are needed in order to understand how rotation impact these constraints and the underlying assumptions. In other words, we can conclude that the spin of M87* is high enough so that the argument based on spherical symmetry cannot be blindly applied to this source.

\newpage
\bibliography{refs}	

\newpage
\revappendix
\section{Tables} \label[appendix]{Tables}
%
\begin{table}[!htbp]
\centering
\small
\begin{ruledtabular}
\begin{tabular}{l@{\qquad}|l|l|l|l|l|l}
$ $ & $\theta = \ang{90}$ & $\theta =\ang{60}$ & $\theta \simeq \ang{47}$ & $\theta =\ang{45}$ & $\theta =\ang{30}$ & $\theta =\ang{1}$\\
\hline
$a \rightarrow 0$ & $1.6875 \cdot 10^{-5} $ &  $1.6875 \cdot 10^{-5} $ &  $1.6875 \cdot 10^{-5} $ &   $1.6875 \cdot 10^{-5} $ & $1.6875 \cdot 10^{-5} $ & $1.6875 \cdot 10^{-5} $\\
$a= 10^{-5}$ & $1.6875 \cdot 10^{-5} $ & $1.6875 \cdot 10^{-5}$ &  $1.6875 \cdot 10^{-5} $ &  $1.6875 \cdot 10^{-5}$ & $1.6875 \cdot 10^{-5}$ & $1.6875 \cdot 10^{-5}$\\
$a= 10^{-3}$ & $1.6875 \cdot 10^{-5} $ & $1.6875 \cdot 10^{-5}$ &  $1.6875 \cdot 10^{-5} $ &  $1.6875 \cdot 10^{-5}$ & $1.6875 \cdot 10^{-5}$ & $1.6875 \cdot 10^{-5}$\\
$a= 0.1$ & $1.6955 \cdot 10^{-5} $ & $1.6920 \cdot 10^{-5}$ &  $1.6890 \cdot 10^{-5} $ &  $1.6885 \cdot 10^{-5}$ & $1.6845 \cdot 10^{-5}$ & $1.6815 \cdot 10^{-5}$\\
$a= 0.5$ & $1.9355 \cdot 10^{-5} $ & $1.8215 \cdot 10^{-5}$ &  $1.7255 \cdot 10^{-5} $& $1.7145 \cdot 10^{-5}$ & $1.6150 \cdot 10^{-5}$ & $1.5245 \cdot 10^{-5}$\\
$a=0.7$ &  $2.3265 \cdot 10^{-5} $ &  $2.0175 \cdot 10^{-5} $ &  $1.7775 \cdot 10^{-5} $ &  $1.7510 \cdot 10^{-5} $ &  $1.5215 \cdot 10^{-5} $ &  $1.3230 \cdot 10^{-5} $\\
$a=0.9$ &  $3.7402 \cdot 10^{-5} $ &  $2.9115 \cdot 10^{-5} $ &  $1.9295 \cdot 10^{-5} $ & $1.8588 \cdot 10^{-5} $ &  $1.3036 \cdot 10^{-5} $ &  $9.1751 \cdot 10^{-6} $\\
$a= 1$ & $2.9204 \cdot 10^{-1}$ & $1.9760 \cdot 10^{-1}$ &  $2.3345 \cdot 10^{-5} $ & $5.3050 \cdot 10^{-7}$ & $7.7200 \cdot 10^{-10}$ & $2.9220 \cdot 10^{-10}$\\
\end{tabular}
\caption{Numerical values of escape probability for $\mu=10^{-5}$.}
\end{ruledtabular}
\end{table}
\begin{table}[!htbp]
\centering
\small
\begin{ruledtabular}
\begin{tabular}{l@{\qquad}|l|l|l|l|l}
$ $ & $\theta = \pi/2$ & $\theta =\pi/3$ & $\theta =\pi/4$ & $\theta =\pi/6$ & $\theta =0.1$\\
\hline
$a \rightarrow 0$ & $1.6875 \cdot 10^{-3} $ &  $1.6875 \cdot 10^{-3} $ & $1.6875 \cdot 10^{-3} $ & $1.6875 \cdot 10^{-3} $ & $1.6875 \cdot 10^{-3} $\\
$a= 10^{-5}$ & $1.685 \cdot 10^{-3} $ & $1.685 \cdot 10^{-3}$ & $1.685 \cdot 10^{-3}$ & $1.685 \cdot 10^{-3}$ & $1.685 \cdot 10^{-3}$\\
$a= 10^{-3}$ & $1.685 \cdot 10^{-3} $ & $1.685 \cdot 10^{-3}$ & $1.685 \cdot 10^{-3}$ & $1.685 \cdot 10^{-3}$ & $1.685 \cdot 10^{-3}$\\
$a= 0.1$ & $1.694 \cdot 10^{-3} $ & $1.690 \cdot 10^{-3}$ & $1.686 \cdot 10^{-3}$ & $1.683 \cdot 10^{-3}$ & $1.680 \cdot 10^{-3}$\\
$a= 0.5$ & $1.930 \cdot 10^{-3} $ & $1.818 \cdot 10^{-3}$ & $1.712 \cdot 10^{-3}$ & $1.612 \cdot 10^{-3}$ & $1.523 \cdot 10^{-3}$\\
$a= 1$ & $2.920 \cdot 10^{-1}$ & $9.880 \cdot 10^{-2}$ & $2.662 \cdot 10^{-3}$ & $7.635 \cdot 10^{-6}$ & $2.917 \cdot 10^{-6}$\\
\end{tabular}
\caption{Numerical values of escape probability for $\mu=10^{-3}$.}
\end{ruledtabular}
\end{table}
\begin{table}[!htbp]
\centering
\small
\begin{ruledtabular}
\begin{tabular}{l@{\qquad}|l|l|l|l|l}
$ $ & $\theta = \pi/2$ & $\theta =\pi/3$ & $\theta =\pi/4$ & $\theta =\pi/6$ & $\theta =0.1$\\
\hline
$a \rightarrow 0$ & $1.6875 \cdot 10^{-1} $ &  $1.6875 \cdot 10^{-1} $ & $1.6875 \cdot 10^{-1} $ & $1.6875 \cdot 10^{-1} $ & $1.6875 \cdot 10^{-1} $\\
$a= 10^{-5}$ & $1.490 \cdot 10^{-1} $ & $1.490 \cdot 10^{-1}$ & $1.490 \cdot 10^{-1}$ & $1.490 \cdot 10^{-1}$ & $1.490 \cdot 10^{-1}$\\
$a= 10^{-3}$ & $1.490 \cdot 10^{-1} $ & $1.490 \cdot 10^{-1}$ & $1.490 \cdot 10^{-1}$ & $1.490\cdot 10^{-1}$ & $1.490 \cdot 10^{-1}$\\
$a= 0.1$ & $1.492 \cdot 10^{-1} $ & $1.490 \cdot 10^{-1}$ & $1.488 \cdot 10^{-1}$ & $1.487 \cdot 10^{-1}$ & $1.485 \cdot 10^{-1}$\\
$a= 0.5$ & $1.550 \cdot 10^{-1} $ & $1.503 \cdot 10^{-1}$ & $1.496 \cdot 10^{-1}$ & $1.491 \cdot 10^{-1}$ & $1.366 \cdot 10^{-1}$\\
$a= 1$ & $3.146 \cdot 10^{-1}$ & $2.453 \cdot 10^{-1}$ & $1.654 \cdot 10^{-1}$ & $1.728 \cdot 10^{-2}$ & $2.381 \cdot 10^{-2}$\\
\end{tabular}
\caption{Numerical values of escape probability for $\mu=10^{-1}$.}
\end{ruledtabular}
\end{table}
 \begin{table}[!htbp]
 \centering
 \small
 \begin{ruledtabular}
\begin{tabular}{l@{\qquad}|l|l|l}
 & Total Probability of photon escape & Errors & Percentage\\
& from surfaces at fixed $r=r_\star$ & &  errors\\
\hline
$a \rightarrow 0$ & $1.6875 \cdot 10^{-5} $ & $0$ & $0$\\
$a=10^{-5}$ & $1.687478 \cdot 10^{-5} $ & $3.9 \cdot 10^{-10}$ & $2.3 \cdot 10^{-3}\ \%$\\
$a=10^{-3}$ & $1.687478 \cdot 10^{-5} $ & $3.9 \cdot 10^{-10}$ & $2.3 \cdot 10^{-3}\ \%$\\
$a=0.1$ & $1.690860 \cdot 10^{-5} $ & $3.9 \cdot 10^{-10}$ & $2.3 \cdot 10^{-3}\ \%$ \\
$a=0.3$ & $1.719878 \cdot 10^{-5}$ & $3.8 \cdot 10^{-10}$ & $2.2 \cdot 10^{-3}\ \%$\\
$a=0.5$ & $1.791177 \cdot 10^{-5} $ & $3.5 \cdot 10^{-10}$ & $2.0 \cdot 10^{-3}\ \%$\\
$a=0.7$ & $1.955220 \cdot 10^{-5}$ & $3.0  \cdot 10^{-10}$ & $1.6 \cdot 10^{-3}\ \%$\\
$a=0.9$ & $2.547805 \cdot 10^{-5} $ & $2.1 \cdot 10^{-10}$ & $8.2 \cdot 10^{-4}\ \%$\\
$a=0.998$ &$1.445679 \cdot 10^{-4}$ &$4.0  \cdot 10^{-11}$ &$2.8 \cdot 10^{-5}\ \%$\\
$a=1$ & $1.577045255454194 \cdot 10^{-1}$ & $1.1 \cdot 10^{-15}$ & $6.7 \cdot 10^{-13}\ \%$\\
\end{tabular}
\label{tab: AA5}
\caption{Numerical values of total probability of photon escape from the surface at fixed $r=r_\star$ for $\mu=10^{-5}$, with corresponding uncertainties.}
\end{ruledtabular}
\end{table}
  \begin{table}[!htbp]
\centering
\small
 \begin{ruledtabular}
\begin{tabular}{l@{\qquad}|l|l|l}
 & Total Probability of photon escape & Errors & Percentage\\
& from surfaces at fixed $r=r_\star$ & &  errors\\
\hline
$a \rightarrow 0$ & $1.6875 \cdot 10^{-3} $ & $0$ & $0$\\
$a=10^{-5}$ & $1.685287 \cdot 10^{-3} $ & $3.9 \cdot 10^{-8}$ & $2.3 \cdot 10^{-3}\ \%$\\
$a=10^{-3}$ & $1.685288 \cdot 10^{-3} $ & $3.9 \cdot 10^{-8}$ & $2.3 \cdot 10^{-3}\ \%$\\
$a=0.1$ & $1.688614 \cdot 10^{-3} $ & $3.9 \cdot 10^{-8}$ & $2.3 \cdot 10^{-3} \%$ \\
$a=0.3$ & $1.717148 \cdot 10^{-3}$ & $3.8 \cdot 10^{-8}$ & $2.2 \cdot 10^{-3} \%$\\
$a=0.5$ & $1.787247 \cdot 10^{-3} $ & $3.5 \cdot 10^{-8}$ & $2.0 \cdot 10^{-3} \ \%$\\
$a=0.7$ & $1.948401 \cdot 10^{-3}$ & $3.0 \cdot 10^{-8}$ & $1.6 \cdot 10^{-3} \ \%$\\
$a=0.9$ & $2.528229 \cdot 10^{-3} $ & $2.1 \cdot 10^{-8}$ & $8.4 \cdot 10^{-4}\ \%$\\
$a=0.998$ &$1.275911 \cdot 10^{-2}$ &$4.0 \cdot 10^{-9}$ &$3.2 \cdot 10^{-5}\ \%$\\
$a=1$ & $1.56531384886 \cdot 10^{-1}$ & $1.1 \cdot 10^{-11}$ & $6.7\cdot 10^{-9}\ \%$\\
\end{tabular}
\label{tab: AA3}
\caption {Numerical values of total probability of photon escape from the surface at fixed $r=r_\star$ for $\mu=10^{-3}$, with corresponding uncertainties.}
\end{ruledtabular}
\end{table}
\begin{table}[!htbp]
\begin{ruledtabular}
\centering
\small
\begin{tabular}{l@{\qquad}|l|l|l}
 & Total Probability of photon escape & Errors & Percentage\\
& from surfaces at fixed $r=r_\star$ & &  errors\\
\hline
$a \rightarrow 0$ & $1.6875 \cdot 10^{-1}$ & $0$ & $0$\\
$a=10^{-5}$ & $ 1.489792 \cdot 10^{-1}$ & $3.4 \cdot 10^{-6}$ & $0.0023\ \%$\\
$a=10^{-3}$ & $1.489792 \cdot 10^{-1}$ & $3.4 \cdot 10^{-6}$ & $0.0023\ \%$\\
$a=0.1$ & $1.489587 \cdot 10^{-1}$ & $3.4 \cdot 10^{-6}$ & $0.0023\ \%$ \\
$a=0.3$ & $1.488248 \cdot 10^{-1}$ & $3.3 \cdot 10^{-6}$ & $0.0022\ \%$\\
$a=0.5$ & $1.487566 \cdot 10^{-1} $ & $3.3 \cdot 10^{-6}$ & $0.0021\ \%$\\
$a=0.7$ & $1.494987 \cdot 10^{-1}$ & $2.8 \cdot 10^{-6}$ & $0.0018\ \%$\\
$a=0.86$ & $1.529864 \cdot 10^{-1}$ & $2.3 \cdot 10^{-6}$ & $0.0014\ \%$\\
$a=0.87$ & $1.534660 \cdot 10^{-1}$ & $2.2 \cdot 10^{-6}$ & $0.0014\ \%$\\
\end{tabular}
\label{tab: AA1}
\caption {Numerical values of total probability of photon escape from the surface at fixed $r=r_\star$ for $\mu=10^{-1}$, with corresponding uncertainties.}
\end{ruledtabular}
\end{table}
%

\end{document}